\documentclass[twocolumn,dvipsnames]{aastex62}
\usepackage{amsmath,amstext}
\usepackage{apjfonts}
\usepackage[figure,figure*]{hypcap}
\usepackage{ulem}
\usepackage{xspace}
\usepackage{xcolor}
\usepackage{lineno}
\usepackage{xfrac}
\usepackage{graphicx}
\usepackage{natbib}
\usepackage{multimedia}
\usepackage{amssymb}
\usepackage{amsmath}
\usepackage{array,booktabs}
\usepackage{mathptmx}
\usepackage{booktabs}
\usepackage{hyperref}
\usepackage{float}
\usepackage{todonotes}
\usepackage{multirow}
\newcommand{\mbh}{\,{M_{\rm BH}}}
\newcommand{\abh}{\,{a_{\rm BH}}}
\newcommand{\erg}{\,{{\rm erg}}}
\newcommand{\G}{\,{{\rm G}}}
\newcommand{\msun}{\,{M_{\odot}}}
\newcommand{\RNS}{\,{{R_{\rm NS}}}}
\newcommand{\RBH}{\,{{R_{\rm BH}}}}
\newcommand{\tBH}{\,{{t_{\rm BH}}}}
\newcommand{\tspin}{\,{{t_{\rm spin}}}}
\newcommand{\tB}{\,{{t_{\rm B}}}}
\newcommand{\s}{\,{{\rm s}}}
\newcommand{\ms}{\,{{\rm ms}}}
\newcommand{\km}{\,{{\rm km}}}
\newcommand{\cm}{\,{{\rm cm}}}
\newcommand{\rg}{\,{r_{\rm g}}}

\definecolor{jaredcolor}{HTML}{5D3FD3}

\shorttitle{She’s Got Her Mother’s Hair: Unveiling the Origin of Black Hole Magnetic Fields}
\shortauthors{Gottlieb, Renzo, Metzger, Goldberg \& Cantiello}

\begin{document}



\title{She’s Got Her Mother’s Hair: Unveiling the Origin of Black Hole Magnetic Fields through Stellar to Collapsar Simulations}

    \author[0000-0003-3115-2456]{Ore Gottlieb}
	\email{ogottlieb@flatironinstitute.org}
    \affiliation{Center for Computational Astrophysics, Flatiron Institute, 162 5th Avenue, New York, NY 10010, USA}
    \affil{Department of Physics and Columbia Astrophysics Laboratory, Columbia University, Pupin Hall, New York, NY 10027, USA}

    \author[0000-0002-6718-9472]{Mathieu Renzo}
    \affil{University of Arizona, Department of Astronomy \& Steward Observatory, 933 N. Cherry Ave., Tucson, AZ 85721, USA}

    \author[0000-0002-3635-5677]{Brian D. Metzger}
    \affiliation{Department of Physics and Columbia Astrophysics Laboratory, Columbia University, Pupin Hall, New York, NY 10027, USA}
    \affiliation{Center for Computational Astrophysics, Flatiron Institute, 162 5th Avenue, New York, NY 10010, USA}

    \author[0000-0003-1012-3031]{Jared A. Goldberg}
    \affiliation{Center for Computational Astrophysics, Flatiron Institute, 162 5th Avenue, New York, NY 10010, USA}

    \author[0000-0002-8171-8596]{Matteo Cantiello}
    \affiliation{Center for Computational Astrophysics, Flatiron Institute, 162 5th Avenue, New York, NY 10010, USA}
    \affiliation{Department of Astrophysical Sciences, Princeton University, Princeton, NJ 08544, USA}

\begin{abstract}
    Relativistic jets from a Kerr black hole (BH) following the core collapse of a massive star (``collapsar'') is a leading model for gamma-ray bursts (GRBs). However, the two key ingredients for a Blandford-Znajek powered jet $-$ rapid rotation and a strong magnetic field $-$ seem mutually exclusive. Strong fields in the progenitor star's core transport angular momentum outwards more quickly, slowing down the core before collapse. Through innovative multidisciplinary modeling, we first use MESA stellar evolution models followed to core collapse, to explicitly show that the small length-scale of the instabilities -- likely responsible for angular momentum transport in the core (e.g., Tayler-Spruit) -- results in a low {\it net} magnetic flux fed to the BH horizon, far too small to power GRB jets. Instead, we propose a novel scenario in which collapsar BHs acquire their magnetic ``hair'' from their progenitor proto-neutron star (PNS), which is likely highly magnetized from an internal dynamo. We evaluate the conditions for the BH accretion disk to pin the PNS magnetosphere to its horizon immediately after the collapse. Our results show that the PNS spin-down energy released before collapse matches the kinetic energy of Type Ic-BL supernovae, while the nascent BH's spin and magnetic flux produce jets consistent with observed GRB characteristics. We map our MESA models to 3D general-relativistic magnetohydrodynamic simulations and confirm that accretion disks confine the strong magnetic flux initiated near a rotating BH, enabling the launch of successful GRB jets, whereas a slower-spinning BH or one without a disk fails to do so.
\end{abstract}

\section{Introduction}

Massive star collapse constitutes the primary formation channel for
neutron stars (NSs) and black holes (BHs) in the Universe (e.g.,
\citealt{Janka12,Burrows&Vartanyan21, soker:24}). Observations of distant
supernovae (SNe) and their remnants in our Galaxy indicate that a
significant fraction of NSs are born as ``magnetars,'' with dipole
magnetic field strengths of $ B \approx 10^{14}-10^{15}\,\G $
\citep{Kaspi2017,Beniamini2019}, some of which may be rapidly spinning
millisecond (ms) magnetars \citep{Kasen2010,Mosta2015}. Based on the
light curves of stripped-envelope SNe (SESNe), it has been argued that
all SESNe, and possibly all SNe, give birth to either a millisecond (ms) magnetar or
an accreting BH \citep{Woosley+21,Rodriguez2024}.

In a small fraction of collapsing stars, the newly formed compact object powers a magnetized relativistic jet capable of generating a gamma-ray burst (GRB). Whether the electromagnetically-driven jets are launched by a BH (\citealt{Blandford1977}, hereafter BZ; \citealt{Popham1999,MacFadyen1999}) or an NS \citep[][see however \citealt{Mosta2014}]{Goldreich1969, Usov1992,Thompson1994,Metzger2007,Metzger2011}, they necessitate rapid rotation and an extremely strong large-scale poloidal magnetic field $\gtrsim 10^{15}$ G to facilitate jet launching (e.g., \citealt{Tchekhovskoy&Giannios15,Gottlieb2022a}). This raises the question of whether a rapidly rotating star, whose core prior to collapse is predicted to possess a comparatively weak magnetic field $B \approx 10^{10}\,\G $ \citep{Aguilera-Dena2018,Muller2024}, can generate such strongly magnetized compact objects.

Two possible magnetar formation channels have been proposed (e.g., \citealt{Spruit2008,Igoshev2021} for reviews). In the ``fossil field'' scenario, an NS acquires its magnetic field externally, from the collapsing star \citep{Ferrario2006}. If the field has a substantial poloidal component, magnetic flux freezing during the collapse will amplify the field by a factor of $(R_c/\RNS)^2$, where $R_c$ is the radius of the precollapse star enclosing the NS mass and $\RNS$ is the NS radius. This factor can be $\gtrsim 10^5$, sufficient to generate an NS magnetic field $\gtrsim 10^{15}\,\G$ \citep[see e.g.,][]{Makarenko2021}. However, the topology of the core's magnetic field as predicted by, e.g., the Tayler-Spruit dynamo \citep[TSD;][]{Tayler1973,Spruit2002,Fuller2019a,Petitdemange2023,Skoutnev2024} is primarily toroidal, such that only a small fraction of the total progenitor flux will contribute to the dipole field of the newly formed proto-neutron star (PNS). Furthermore, even if the poloidal field is relatively strong, the small spatial scales that characterize the TSD render the field unlikely to possess a uniform polarity prior to or after the collapse, resulting in a reduced {\it net} flux reaching small scales.

Alternatively, if a small-scale dynamo similar to the TSD indeed dominates angular momentum transport in massive stars \citep[see however, e.g.,][]{kissin:15, denhartogh:20}, then magnetars must acquire their large-scale fields during their formation, i.e. after the collapse phase. A natural source of free energy to amplify the magnetic field in such a scenario is differential rotation within the PNS or an accretion disk surrounding it.

At least three mechanisms can be at play over the first few seconds of the PNS cooling evolution: (1) the magnetorotational instability \citep[MRI;][]{Balbus1991}, particularly in the outer layers of the PNS where the angular velocity decreases outward \citep{Akiyama2003,Obergaulinger2009,Reboul-Salze2021}; (2) the TSD, particularly in stably stratified regions close to the PNS core where the MRI cannot operate (e.g., \citealt{Margalit2022}); and (3) an $\alpha-\omega$ dynamo in the convective regions of the PNS \citep[e.g.,][]{Duncan1992,Thompson1993,Bonanno2003,Raynaud2020,White2022}. All three mechanisms are in principle capable of generating fields of strength $\lesssim 10^{16}\,\G$ \citep{Thompson2001,Akiyama2003}, though only the MRI and a convective dynamo are likely to dominate near the surface of the PNS (however, see \citealt{Barrere2022}). The Rossby number $Ro = P/\tau_c$ determines the relative contribution of these mechanisms, where $P$ is the PNS spin period and $\tau_c$ is the convective turnover time in its envelope. When rotation is slower than the convective motions such that $Ro \gtrsim 1$, the MRI dominates as it grows on the rotation timescale, leading to a stochastic dynamo that generates small-scale multipolar fields (e.g., \citealt{Guilet+22,Jacquemin-Ide+23}). On the other hand, for sufficiently rapid rotation $P \lesssim 10$ ms and $ Ro \lesssim 0.2 $, the MRI is subdominant to the $\alpha-\omega$ dynamo, which can generate a strong ordered dipolar field \citep{Nagakura2020,Raynaud2020,White2022}, whose magnitude grows with the rotation rate \citep{Masada2022}. 

While the origin of the NS magnetic fields has been extensively debated in the literature, the question of how BHs acquire enough magnetic flux to power relativistic jets in jetted core-collapse events like GRBs has received less attention. Unlike NSs, BHs cannot generate their own fields, which must be inherited externally. One possibility is that the BH acquires its magnetic flux directly through accretion from the infalling progenitor envelope (e.g., \citealt{Tchekhovskoy&Giannios15}). However, as we will show, the same arguments raised above against magnetars acquiring large-scale fields from the progenitor star -- namely the strength and small coherence scale of the precollapse poloidal field -- also apply to the accreting BH case, particularly for stellar cores that retain enough angular momentum to produce GRB jets in the first place. Another possibility is an amplification of the field through dynamo processes in the accretion disk \citep[e.g.,][]{Jacquemin-Ide+23,Most2023}. This mechanism might also be inefficient for collapsar GRBs given the expected properties of collapsar disks, as we discuss later in light of our stellar evolution models. On the other hand, stellar-mass BHs were previously PNSs, prior to accreting enough mass to undergo gravitational collapse (e.g., \citealt{Dessart+12,Aloy2021}). This motivates the possibility that collapsar BHs {\it inherit their magnetic fields from their progenitor PNS.}

Simulations of the gravitational collapse of an isolated NS into a BH \citep{Baumgarte2003,Lehner2012,Dionysopoulou2013,Nathanail2017,Most2024} show that the BH can at least initially inherit the (external) magnetic field of the NS, albeit with some rearrangement of the magnetosphere geometry. In particular, flux freezing from the NS surface to the more compact BH horizon increases the poloidal magnetic field strength by a factor of $(\RNS/\RBH)^2 \approx 10$, with a sizable toroidal field component generated by differential rotation in the magnetosphere. However, in the absence of an external source of plasma, such a magnetosphere is short-lived. Current sheets quickly form along oppositely directed field lines, creating buoyantly rising plasmoids, which untether and carry away the BH magnetosphere, ultimately in the form of a large-scale electromagnetic shock wave (e.g., \citealt{Most2024}). As a result, the BH loses its magnetic flux over tens to hundreds of horizon light crossing times (ms timescales; e.g., \citealt{Bransgrove+21}). The ultimate loss of magnetic flux is an inevitable consequence of the BH ``no-hair theorem'' in the presence of magnetic dissipation.

However, this ``balding'' process may be postponed if the inherited magnetic field can be pinned to the BH by an external current source, such as that supplied by an accretion disk. While not expected to form around an old NS that collapses in isolation (e.g., accretion-induced collapse in a compact binary; \citealt{Margalit2015}), such a disk may be present around a differentially rotating PNS embedded within the infalling envelope of a rotating star. Nevertheless, it remains unaddressed whether, in astrophysical systems such as collapsars or post-binary NS mergers, the BH can maintain the inherited magnetic flux long enough to launch a sustained relativistic jet necessary to power GRBs.

\begin{figure*}[]
  \centering
  \includegraphics[width=1\textwidth]{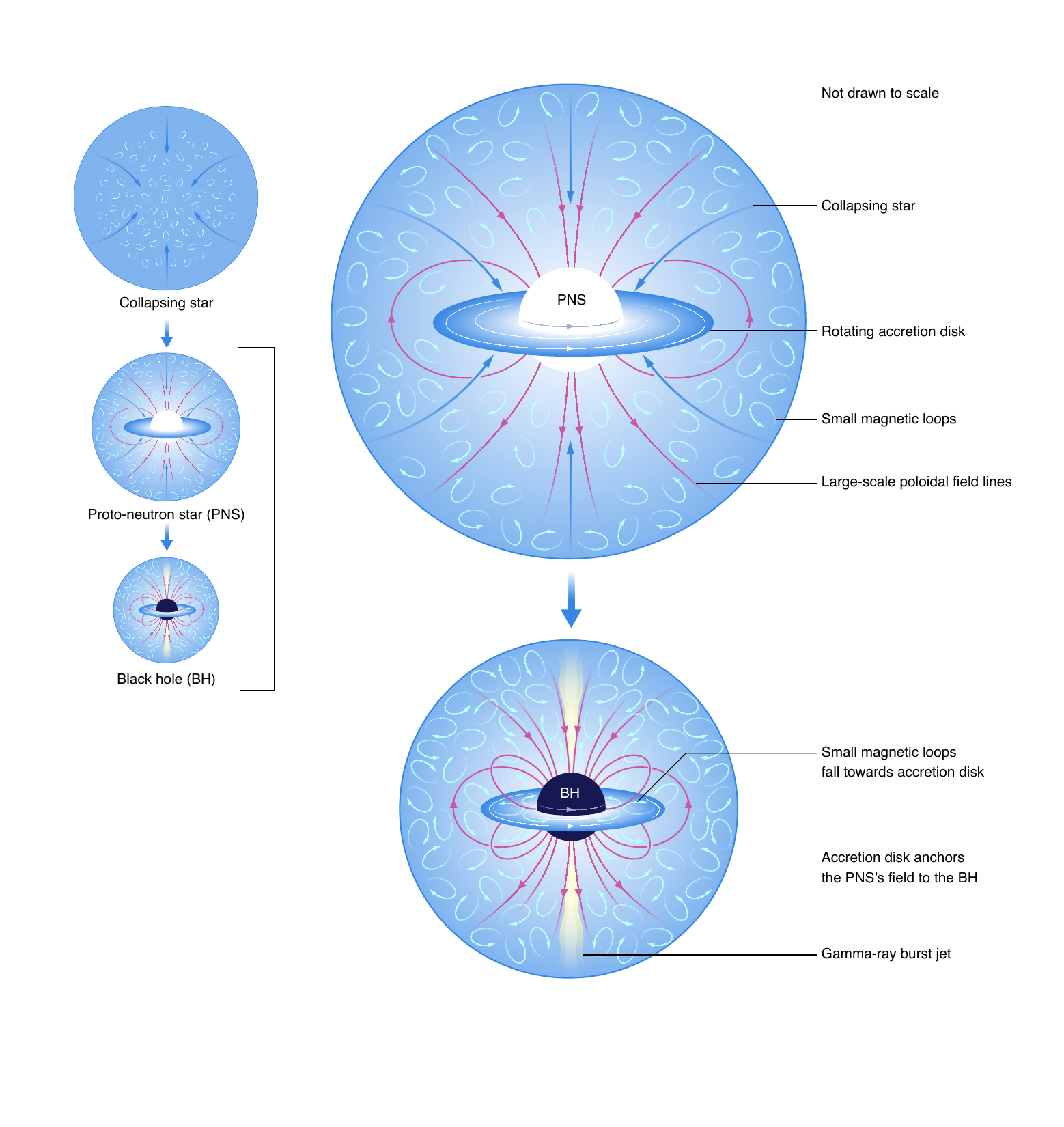}
  \caption{Not-to-scale illustration of the proposed model. Top: a rotating massive star (light blue) collapses, forming a rapidly rotating proto-neutron star (white) surrounded by an accretion disk (dark blue) due to the excess angular momentum of the innermost layers of the stellar core (\S\ref{sec:remnant}). The magnetic field present in the progenitor star due to the Tayler-Spruit dynamo is too weak and/or small-scale (randomly oriented cyan loops) to contribute a substantial magnetic flux onto the compact object (\S\ref{sec:flux}). A convective dynamo within the rapidly spinning proto-neutron star efficiently generates a strong, ordered magnetic field (pink lines).
  Bottom: the proto-neutron star collapses into a moderately spinning black hole (black). The accretion disk surrounding the proto-neutron star at the time of collapse quickly spreads inwards viscously to the black hole's innermost stable circular orbit, anchoring the proto-neutron star's magnetic field lines to the BH faster than the ``balding'' process. This enables the black hole to form with strong magnetic fields already threading through it (\S\ref{sec:bh}). The strongly magnetized and moderately spinning black hole then launches a pair of Blandford-Znajek jets (yellow) that break through the outer layers of the collapsing star and power gamma-ray burst emission (\S\ref{sec:grmhd}).
  }
  \label{fig:sketch}
\end{figure*}

Here, using an innovative multidisciplinary approach, we explicitly show that massive stars with rapidly rotating cores at the time of collapse generate an accretion disk around the central compact object but do not provide sufficient magnetic flux to power a GRB-like jet. Instead, we propose that such events first create a rapidly spinning PNS surrounded by an accretion disk and that internal dynamo processes amplify the PNS magnetic field to magnetar-level strengths. Then, when the magnetar accretes sufficient mass to collapse, its strong poloidal component is retained by the resulting BH due to confinement by the accretion disk. Therefore, even without ongoing accreted magnetic flux, the BH retains its dipolar field, which can launch the BZ jets that generate GRBs. The astrophysical setup and sequence of events are illustrated in Figure~\ref{fig:sketch}.

The paper is organized as follows. In \S\ref{sec:remnant}, we begin by presenting stellar evolution models that lead to the formation of a PNS with an accretion disk and use them to explain why the intrinsic poloidal flux fed by stellar infall is likely insufficient to power a GRB jet. In \S\ref{sec:bh}, we estimate the conditions necessary for the BH to capture the field of the collapsing PNS, which constrain the expected natal spin and magnetic flux acquired by the BH during this process. In \S\ref{sec:grmhd}, we conduct general-relativistic magnetohydrodynamic (GRMHD) simulations demonstrating that BHs retain their magnetic field without further flux accumulation so long as an accretion disk confines them. We conclude in \S\ref{sec:discussion}, and discuss the implications of our results on GRBs and their associated SNe in \S\ref{sec:implications}.

\section{Black hole formation in stellar evolution models}\label{sec:remnant}

We compute a stellar progenitor using MESA \citep{Paxton2011,Paxton2013,Paxton2015,Paxton2018,Paxton2019,Jermyn2023}, starting with a metallicity of $Z=0.001$ and fast initial rigid rotation corresponding to $\omega=0.6\,\omega_\mathrm{crit}$, where $\omega_\mathrm{crit}=\sqrt{(1-L/L_\mathrm{Edd})GM/R^3}$ is the surface critical rotation rate accounting for radiative forces, $L$ is the stellar luminosity, $L_\mathrm{Edd}$ the Eddington luminosity, $G$ is the gravitational constant, $M$ is the total mass (initially 40$\,M_\odot$) and $R$ the stellar radius. Such an extremely high initial rotation rate is unseen in the local Universe \citep[e.g.,][]{ramirez-agudelo:15}, but guarantees rotationally-induced chemically homogeneous evolution \citep{maeder:00,Yoon:2005,Woosley:2006,Yoon:2006,Cantiello:2007}. We discuss the details of this progenitor's evolution in Appendix~\ref{appendix:MESA} and provide input and output files at \href{https://doi.org/10.5281/zenodo.12193630}{doi.org/10.5281/zenodo.12193630}.

\begin{figure}[bp]
  \centering
  \includegraphics[width=0.5\textwidth]{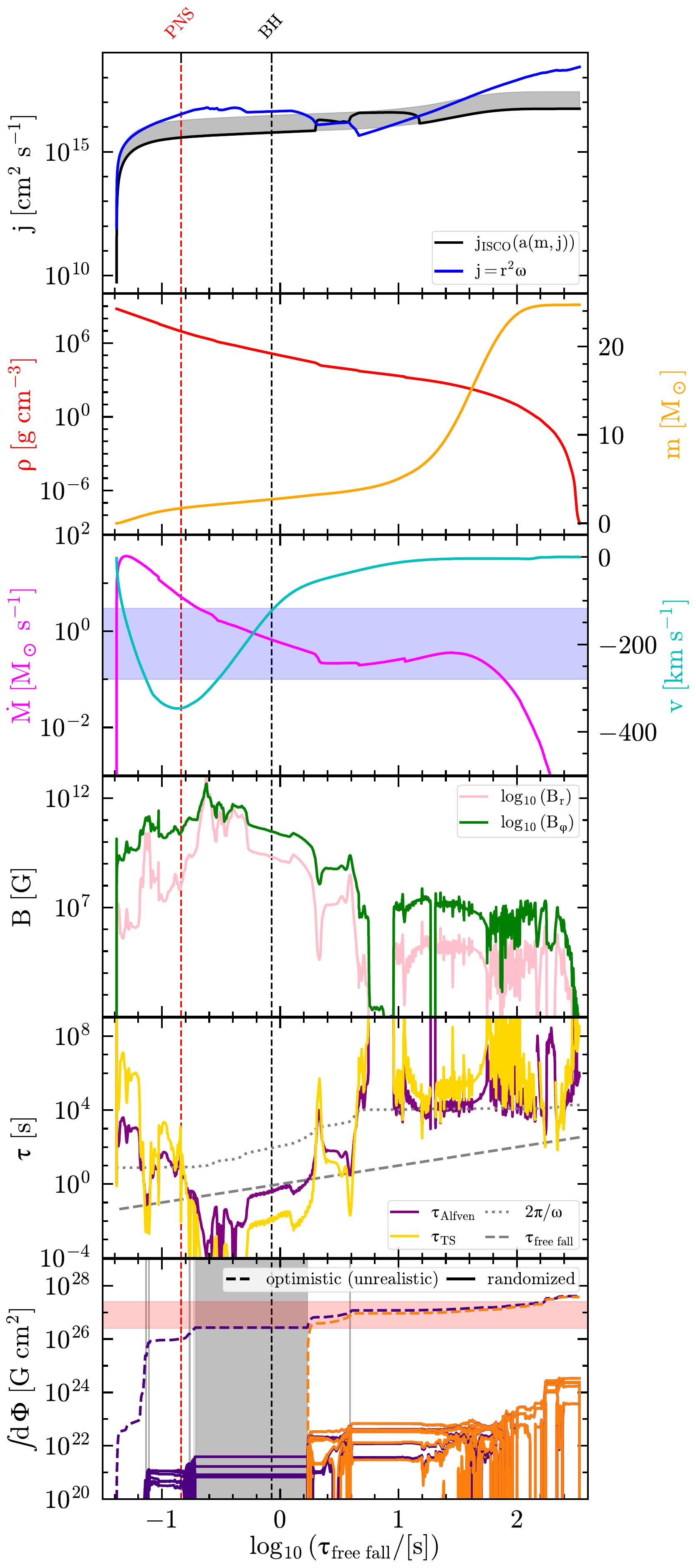}
  \caption{From top to bottom, as a function of the local free fall timescale: specific angular momentum profile (gray band brackets $j_\mathrm{ISCO}(a_\mathrm{BH}=0)$ and $j_\mathrm{ISCO}(a_\mathrm{BH}=1)$); mass density (red) and enclosed mass (orange; right vertical axis); estimate of the accretion rate (magenta) and local infall velocity profile (cyan; right vertical axis); B-field components predicted by the Tayler-Spruit dynamo; Alfv\'{e}n, TSD, and free fall local timescales (purple, yellow, and dashed gray, respectively); and estimated cumulative magnetic flux $\Phi$, where the red band is the required flux to power a GRB jet (see text).}
  \label{fig:CC_profile}
\end{figure}

Figure~\ref{fig:CC_profile} illustrates the pre-collapse properties as a function of the local free-fall timescale, estimated as:
\begin{equation}
  \label{eq:t_ff}
  \tau_\mathrm{free \ fall}(r) = \sqrt{\frac{3\pi r^3}{32 G m}}\,,
\end{equation}
where $r$ is the radius and $m\equiv m(r)$ is the enclosed mass. The choice of the numerical coefficient in Eq.~\eqref{eq:t_ff} introduces a factor of $\sim10-100$ uncertainty; we adopt a value that yields good agreement between the estimated mass accretion rates (see below) and the results in \cite{Gottlieb2022a}. The vertical dashed lines throughout all panels mark the mass coordinates corresponding to $m=1.7\,M_\odot$, estimating a massive PNS (red), and $m=2.7\,M_\odot$, estimating the mass of a seed BH (black).

The top panel of Fig.~\ref{fig:CC_profile} depicts the specific
angular momentum of the infalling matter, $j= r\omega^2$ (blue), and
the innermost stable circular orbit (ISCO) specific angular momentum,
$j_\mathrm{ISCO}$ (black), as a function of the enclosed mass and angular
momentum calculated following \cite{bardeen:72}. In the innermost
layers, where $\tau_\mathrm{free\ fall}\lesssim 3\,\mathrm{s}$, we
find $j\gtrsim j_\mathrm{ISCO}$, which prohibits a prompt collapse to
a BH and necessitates a phase of a PNS with an accretion disk
\citep[see also][]{Dessart2008}. This result is sensitive to
uncertain angular momentum transport processes in the star, which are
virtually unconstrained in this mass regime. As outlined in
Appendix~\ref{appendix:MESA}, we adopt angular momentum transport via TSD in radiative layers
\citep{Spruit2002}, which provides relatively weak transport by
magnetic torques (compared to \citealt{Fuller2019a,fuller:22}). In convective layers, angular momentum transport is assumed to be quite efficient, with a diffusion coefficient proportional to the convective diffusivity. We also neglect possible late spin-up of the core due to
gravity waves excited by convective shell burning \citep[e.g.][]{fuller:15}.

Using the same initial conditions and incorporating magnetic torques from\footnote{Implemented in \url{https://github.com/MESAHub/mesa-contrib/}.} \cite{fuller:22}, the model still exhibits chemically homogeneous evolution. However, most of the angular momentum is lost post-hydrogen core burning, and no layer retains enough angular momentum at the onset of collapse to form an accretion disk. Different assumptions can result in the inner regions spinning less rapidly, allowing a prompt collapse to a BH \citep{MacFadyen1999}. In such cases, one can estimate the BH seed formation time $t_\mathrm{BH}$ as the free fall time of the layers with $j<j_\mathrm{ISCO}(a(m, j))$, and the accretion disk might form from the outer parts of the stellar envelope. This raises difficulties for this scenario in producing GRB progenitors.

The second panel portrays the mass density profile (red; left vertical axis), and the enclosed mass (orange; right vertical axis). The total mass of the star at the onset of core collapse is $\sim{}25\,M_\odot$ due to wind mass loss (see Appendix~\ref{appendix:MESA}). The third panel shows the infall velocity profile (cyan; right axis;=); we evolve the star until it reaches an infall velocity of $-300\,\mathrm{km\ s^{-1}}$. The third panel also shows a rough estimate of the mass accretion rate calculated as $\dot{M} = {\rm d}m/{\rm d}\tau_\mathrm{free\ fall}$ (magenta). The horizontal
blue shaded band highlights $0.1\leq \dot{M}/M_\odot\,\s^{-1} \leq 3$ (see \S\ref{sec:bh}).

The fourth panel depicts the radial ($ B_r $, pink) and azimuthal ($ B_\varphi $, green) components of the magnetic field $B$ predicted by the TSD \citep{Spruit2002}. The region where the field drops below 100\,G is convective at the onset of core collapse and thus does not contain TSD-generated fields. These fields are, by definition, local and do not maintain the large-scale structure or a poloidal topology, both of which are required to power electromagnetically driven jets (see \S\ref{sec:flux}).

The fifth panel delineates the local Alfv\'{e}n timescale (purple):
\begin{equation}
  \label{eq:Alfven}
  \tau_\mathrm{Alfven} = \frac{\ell}{v_\mathrm{Alfven}} = \frac{\ell}{B_r/(4\pi \rho)^{1/2}} \,,
\end{equation}
where $\ell = 1/ k_{c}$ is the largest radial TSD length-scale (see Eq.~10 in \citealt{Spruit2002} for the critical wavelength $k_{c}$), and $\rho$ is the mass density. The timescale for the growth of B-fields via TSD in radiative layers is (yellow):
\begin{equation}
  \label{eq:tau_TS}
  \tau_\mathrm{TS} = \tau_\mathrm{Alfven}^{2} \times \omega \,,
\end{equation}
where $\omega$ is the local rotational frequency
\citep[e.g.][]{pitts:85, spruit:99}. This implicitly assumes
$\tau_\mathrm{Alfven}\gg 2\pi/\omega$. We show the rotational period
$2\pi/\omega$ as a dotted gray line in the fifth panel, which
illustrates that this assumption holds within the mass that will form the PNS (red vertical dashed line) and in the outermost layers with free-fall times longer than $\simeq 1\,\mathrm{sec}$. However, in between, the assumption of flux freezing does not hold. 
In the innermost layers, with $m\lesssim M_\odot$ and
$\tau_\mathrm{free \ fall}\lesssim 0.1\,\mathrm{s}$, both
$\tau_\mathrm{Alfven}$ and $\tau_{TS}$ are significantly longer than
the free-fall timescale, justifying the assumption of flux freezing.
Moving outward in the region $\msun \lesssim m \lesssim 3\,\msun$,
corresponding to infall times
$\tau_\mathrm{free \ fall}\lesssim 2\,\mathrm{s}$, the Alfv\'{e}n and
TSD timescales become shorter than the infall timescale. Thus,
magnetic diffusion and reconnection could occur, leading to a
lower magnetic flux reaching the seed BH (without dynamos).

\subsection{Advected flux on the horizon}\label{sec:flux}

The last panel of Figure~\ref{fig:CC_profile} shows the cumulative magnetic flux $\Phi = \int {\rm d}\Phi$ reaching the forming compact object from the local B-fields generated by the TSD during stellar evolution. We consider two approaches to estimate this:

\begin{enumerate}
\item The magnetic field lines in different regions of the collapsing
  star likely have random polarity, causing their contribution to the
  flux to at least partly cancel out during accretion. Nevertheless,
  for the sake of illustration, we adopt an unrealistically optimistic
  approach in which the \emph{local} components of the magnetic flux
  are assumed to add up coherently as:
  \begin{equation}
    \label{eq:unrealistic_dB}
    {\rm d}\Phi = 2\pi r B_{r} {\rm d}r\,.
  \end{equation}
\item Consider that the TSD acts to generate local ``loops'' of a
  roughly uniform magnetic field with a linear dimension
  $\lesssim \ell$ \citep{Spruit2002}. Each shell of the star at radius
  $r$ and thickness $\ell$ then contains
  $N_\mathrm{loops} = 4\pi r^2\ell/ (4 \pi \ell^3/3)$ loops, each with
  a random polarity. Assuming $N_\mathrm{loops}\gg1$, we expect a
  random walk in the polarity of the loops, resulting in a total
  contribution for a shell of radius $r$ and thickness $\ell$ equal to
  $\pm {N^{-1/2}_\mathrm{loops}}$ of the right-hand side of
  Eq.~\eqref{eq:unrealistic_dB}. Stellar evolution codes cannot
  resolve the length-scale $\ell$ throughout the model. For most of
  the region of interest here ($\log_{10}(r/\mathrm{cm})\lesssim 10$),
  we find $\ell\ll {\rm d}r$ where ${\rm d}r$ is the spatial resolution of the
  MESA model. Thus, to obtain the contribution of a MESA shell, we
  average the contributions of all (sub-grid) shells of thickness $\ell$ within a
  shell of thickness ${\rm d}r$. These are
  $N_\mathrm{shells} = {\rm d}r/\ell$, and once again assuming
  $N_\mathrm{shells}\gg1$, and assuming each of these shells to have a
  random polarity, their summed contribution will bring a factor of
  $\pm {N^{-1/2}_\mathrm{shells}}$ to the right-hand side of
  Eq.~\eqref{eq:unrealistic_dB}. Overall, a more realistic estimate of
  the contribution of the differential flux is:
  \begin{equation}
    \label{eq:realistic_dB}
    {\rm d}\Phi = \pm\frac{2\pi r B_{r} {\rm d}r}{\sqrt{N_\mathrm{loops}N_\mathrm{shells}}}
    \equiv \pm \frac{2\pi r B_{r} {\rm d}r}{\sqrt{\frac{3r^2{\rm d}r}{\ell^3}}}\,.
  \end{equation}
  We note that the estimated $N_{\rm shell}$ and $N_{\rm loops}$ are
  not always much larger than 1 throughout the star (especially in the
  outermost layers). In this case, our approximation of a random walk
  in polarity may not hold, but this should not result in a much larger
  integrated $\Phi$.
\end{enumerate}

The bottom panel of Fig.~\ref{fig:CC_profile} compares the
unrealistically optimistic estimate (dashed lines) and the randomized
polarity of TSD magnetic loops [solid lines for 10 random draws of the
sign of each ${\rm d}\Phi$ contribution from Eq.~\eqref{eq:realistic_dB}].
The purple lines start the integration from the center, while the
orange lines assume no contribution from the inner
$m = \mbh = 2.7\,M_\odot$.

Only in the most unrealistic scenario does
the cumulative flux reach $\sim10^{27}\,\mathrm{G\ cm^{2}}$ at
$\tau_\mathrm{free\ fall}\approx 300\,\mathrm{s}$. In the realistic
approach, the integrated magnetic flux reaches merely
$ \Phi \approx 10^{23}\,\mathrm{G\ cm^2}$. In both estimates, we
assume that a layer can contribute to the cumulative magnetic flux
only when $\tau_\mathrm{Alfven}>\tau_\mathrm{free\ fall}$. This
guarantees that the local magnetic fields do not have time to decay
and we can assume flux freezing. When this condition is not met (gray
areas), we assume the magnetic fields produced during the stellar
evolution decay completely during the collapse, contributing zero to
the flux:
${\rm d}\Phi(\tau_\mathrm{Alfven}<\tau_\mathrm{free\ fall}) \equiv 0$.

Once a BH forms and spins up due to disk accretion, it could conceivably power a relativistic jet capable of escaping the star \citep{MacFadyen1999}. The power of the BZ jet from a Kerr BH of mass $\mbh$ depends on both its (dimensionless) spin $\abh$ and the magnetic flux according to \citep{Tchekhovskoy2011,Gottlieb2023e}:
\begin{eqnarray}\label{eq:Pj}
    &P_j& = \dot{M}\eta_\phi\eta_ac^2 \approx \left(\frac{\Phi_{\rm B}}{\phi_{\rm MAD}}\right)^2\frac{\eta_ac}{\rg^2} \nonumber \\
    &\underset{\abh \lesssim 0.5}\approx& 3\times 10^{49}\,{\rm erg\,s^{-1}}\,\left(\frac{\Phi_{\rm B}}{10^{27}\,{\rm G\,cm^{2}}}\right)^{2}\left(\frac{M_{\rm BH}}{3\,\msun}\right)^{-2}\left(\frac{\abh}{0.5}\right)^{2}\,,
\end{eqnarray}
where $ \eta_a = 1.063\abh^4 + 0.395\abh^2 $ \citep{Lowell2024} and $ \eta_\phi = (\Phi/\phi_{\rm MAD})^2(\dot{M}\rg^2c)^{-1} $ are the spin and flux efficiencies, respectively, and $ \rg $ is the BH gravitational radius. 
Here, $\phi_{\rm MAD} \approx 50$ is the maximum dimensionless magnetic flux \citep{Tchekhovskoy2011}, corresponding to a magnetically arrested disk (MAD). Thus, even for a rapidly spinning BH, a magnetic flux $\Phi \gtrsim 10^{27}$ G cm$^{2}$ is required to explain the typical luminosities $L_{\rm j} \gtrsim 10^{49}\,\erg\,\s^{-1} $ of GRB jets (e.g., \citealt{Butler2010,Wanderman2010}). As illustrated by the red shaded region in the bottom panel of Fig.~\ref{fig:CC_profile}, which assumes $\abh=0.5$ and $M_\mathrm{BH}=2.7\,\msun$, only the most unrealistically optimistic scenario in our progenitor model can provide such a flux. In the realistic scenario of canceling polarities, the required magnetic flux falls short by several orders of magnitude. We conclude that the magnetic fields generated in the progenitor star are likely insufficient for the BH to launch a relativistic jet.

One potential caveat is that the TSD produces larger fields closer to the rotation axis \citep{Spruit2002, Skoutnev2024}, where rotational support by centrifugal forces is lower. Both of our calculation approaches assume spherical symmetry when estimating ${\rm d}\Phi$, as does the underlying stellar evolution model. This assumption may lead to a mild underestimation of the magnetic flux reaching the PNS from the collapse along the polar regions.

\section{BH formation}\label{sec:bh}
\subsection{Preserving the PNS's magnetic field}
Having established the progenitor star to be an unlikely source of magnetic
flux, we now consider the conditions under which the nascent BH can
inherit the magnetic field of the PNS. For this to occur, the PNS
magnetosphere must be anchored close to the PNS surface by an
accretion disk prior to and following BH formation. Absent such a
disk, the newly formed BH will quickly shed its magnetic field.
Resistive GRMHD simulations find that this ``balding'' process occurs on a
timescale \citep[e.g.,][]{Bransgrove+21,Selvi2024}:
\begin{equation}
t_{\rm bald} \approx 500\frac{\rg}{c}\,,
\end{equation}
where the large prefactor follows from the slow reconnection rate $v_{\rm rec} \approx 0.01v_{\rm A} \approx 0.01 c$ in ideal (collisional) magnetohydrodynamics (e.g., \citealt{Bhattacharjee+09,Uzdensky2010}).\footnote{The collisionless reconnection rate is closer to $\approx 0.1c$ (e.g., \citealt{Bransgrove+21}), resulting in faster BH balding. However, the plasma densities in collapsars are sufficiently large to be highly collisional.}  Prior to the collapse, the accretion disk surrounding the PNS extends down to the Alfv\'{e}n radius (e.g., \citealt{Alpar2001}):
\begin{equation}\label{eq:RA}
   R_A \approx 3\left[\left(\frac{B}{10^{15}\,{\rm G}}\right)^2\left(\frac{\RNS}{10\,\km}\right)^6\left(\frac{\dot{M}}{\msun\,\s^{-1}}\right)^{-1}\right]^{2/7} \,{\rm km}\,,
\end{equation}
where $\dot{M}$ is the mass accretion rate. After the BH forms, the disk spreads from $\gtrsim R_A$ down to the BH horizon, on a timescale roughly given by the viscous time at $R_A$:
\begin{equation}
    t_{\rm visc} \approx \frac{1}{\alpha_\nu \Omega_k}\left(\frac{h}{r}\right)^{-2}\Bigg|_{R_{A}} \approx 200\left(\frac{0.3}{\alpha_\nu}\right)\left(\frac{h/r}{0.3}\right)^{-2}\left(\frac{R_A}{3\rg}\right)^{3/2}\frac{\rg}{c},
\end{equation}
where $\Omega_k \approx (G\mbh/r^{3})^{1/2}$ is the Keplerian angular frequency, $\alpha_\nu$ is the effective kinematic viscosity which at the PNS surface might be well above $ \alpha_\nu \sim 0.3 $ due to magnetic torques \citep[see e.g.,][]{Manikantan2024}, and $ r$ and $h$ are the horizontal and vertical scale-height of the disk, respectively.

If $R_A \gg \RNS \approx 3\rg$, then after the collapse, the newly untethered magnetic field will expand away from the BH on a timescale much faster than the accretion disk can reach the BH surface from $R_A$ to hold the flux in place, i.e. $t_{\rm visc} \gg t_{\rm bald}.$ Thus, for very strong magnetic fields and/or low-accretion rates, $R_A \gg \RNS$, and the BH is expected to form with little of the PNS's magnetic flux. On the other hand, if $R_A \lesssim \RNS $, then $t_{\rm visc} \lesssim t_{\rm bald}$ and the disk will have time to pin the PNS's flux against the horizon faster than it will be shed.  For much of the parameter space we shall find that $R_A \lesssim \RNS $ ($t_{\rm visc} \lesssim t_{\rm bald}$), implying that BHs are capable of inheriting a significant portion of the PNS's flux.

A centrifugally supported disk is likely to be present around the PNS due to the high angular momentum of the infalling stellar envelope [Fig.~\ref{fig:CC_profile}(a)]. Even if this were not the case, a disk can be created from the PNS itself during the collapse process if the PNS has sufficiently large differential rotation (this is not possible for a cold NS in solid body rotation; \citealt{Margalit2015}). Such a situation is encountered in binary NS mergers, for which numerical simulations \citep{Kiuchi2014,Kawamura2016,Ruiz2017,Ruiz2019,Ruiz2021,Bamber2024} find that even though the strongly magnetized ($B \approx 10^{16}\,\G $) core of the NS remnant is devoured by the BH within a ms after collapse, the high-angular momentum envelope with $B \approx 10^{15}\,\G $ \citep{Palenzuela2022,Aguilera-Miret2023} forms a highly magnetized BH accretion disk, which can feed the BH with magnetic flux.

\subsection{Natal black hole spin and magnetic field}
The maximum magnetic flux a BH can acquire from a PNS reflects that of
the PNS prior to collapse. In the presence of an $ \alpha-\omega $
dynamo (e.g., \citealt{Thompson1993}), the large-scale magnetic field
of the PNS depends on a comparison between its convective overturn
time $ \tau_c $ and rotational period, $P$. For rapidly spinning
progenitor stars of interest, the latter can range from $P \sim \ms $
close to the NS surface to much higher values $ P \gg 1\,\ms $ at greater depths, corresponding to a broad range in the Rossby number throughout the PNS. \citet{Raynaud2020,Aloy2021} found $ Ro \lesssim 1 $ close to the NS surface, supporting an efficient convective dynamo capable of PNS large-scale field amplification by orders of magnitude to $ B \approx 10^{15} $ G.

The spin of the BHs created in the majority of (failed) core-collapse SNe is generally expected to be moderate.
Specifically, spin values of $0.2 \lesssim \abh \lesssim 0.5$ are needed to power GRB jets through the BZ mechanism \citep{Gottlieb2023b}. A PNS rotating at period $P$ at the time of collapse will form a BH of spin:
    \begin{equation}\label{eq:aNS}
    \abh \approx 0.24\left(\frac{\RNS}{12\,\km}\right)^2\frac{1\,\ms}{P}\frac{3\,\msun}{\mbh}~,
    \end{equation}
where $M$ and $ I = 0.35 M\RNS^2 $ are the PNS mass and moment of inertia, the latter is calculated following \citet{Bejger2002} for an assumed compactness of $0.12$ \citep{Dietrich2020}.

After forming but prior to collapse, the PNS rotation rate evolves due to a combination of accretion and magnetic braking. Insofar as the latter only becomes relevant once the magnetic field is amplified, three distinct timescales enter the problem: the PNS collapse time $\tBH $, the magnetic field amplification timescale $\tB$, and the magnetic braking spin-down time, $ \tspin $.

The PNS typically forms with a mass $\approx 1.2-1.6\,\msun$ close to the effective Chandrasekhar mass of the progenitor's iron core and hence must accrete roughly a solar mass to surpass the Tolman-Oppenheimer-Volkoff limit $M_{\rm TOV} \approx 2-2.6\,\msun$ and collapse. The collapse time can thus be estimated as $\tBH \sim \msun/\dot{M}$ and is typically a few seconds for PNS accretion rates $ \dot{M} \sim 0.1-1\,\msun\,\s^{-1}$ (Fig.~\ref{fig:CC_profile}; see also \citealt{Aloy2021}). The magnetic amplification timescale, $\tB$, is also anticipated to be of order seconds \citep{Raynaud2020}, facilitated by the neutrino cooling contraction of the PNS on this timescale \citep{Burrows&Lattimer86} that reduces the Rossby number below the critical threshold \citep{Raynaud2020,Aloy2021}. The fact that $\tBH$ and $\tB$ are comparable implies that the PNS may collapse before or after the amplification of its field has saturated.

When $ \tBH \gtrsim \tB $, the PNS has time to grow a strong magnetic field before collapsing. In this regime, the PNS not only spins up by accreting angular momentum from the disk but also spins down due to the magnetized pulsar-like wind \citep[e.g.,][]{Metzger2007,Metzger2018,Margalit2022}. The PNS angular momentum $J$ evolves under these processes according to\footnote{We neglect secondary effects of accretion of unmagnetized gas that may reduce the PNS dipole field \citep[e.g.,][]{Aloy2021}.} \citep{Metzger2018}:
\begin{equation}\label{eq:spin}
    \dot{J}(t) = \dot{M}\sqrt{GM(t)\RNS}\left(1-\frac{\Omega(t)}{\Omega_k}\right) - \frac{B^2R_{\rm NS}^4\Omega(t)}{c}\,,
\end{equation}
where the factor $ (1-\Omega/\Omega_k) $ limits the PNS to spin below the centrifugal breakup velocity, $\Omega_k$ is now the Keplerian angular velocity at the PNS surface, and
\begin{equation}
    M(t) = M_0 + {\dot{M}}t\, ,\,\,\, t \le t_{\rm BH}\,
\end{equation}
accounts for the growth of the PNS mass due to accretion.
For the magnetic spin-down, we have approximated the magnetic field geometry of the wind as that of a split-monopole, as is appropriate because the polar magnetic field lines are torn open by neutrino-driven mass-loading (e.g., \citealt{Metzger2007,Metzger2011}) and magnetosphere compression due to accretion when $R_A \le \RNS $ (e.g., \citealt{Parfrey+2016,Metzger2018}).

Given sufficient time, an equilibrium between magnetic braking and accretion spin-up is achieved ($\dot{J} = 0$), at an equilibrium spin period
\begin{align}
P_{\rm eq} \approx 1.2\left(\frac{M}{2\,\msun}\right)^{-1/2}\Bigg[&\left(\frac{B}{10^{16}\,{\rm G}}\right)^{2}\left(\frac{\dot{M}}{\msun\,{\rm s}^{-1}}\right)^{-1}\left(\frac{\RNS}{12\,{\rm km}}\right)^{7/2}
\notag\\&
    +0.42\left(\frac{\RNS}{12\,{\rm km}}\right)^{3/2}\Bigg]\,{\rm ms}\,,
\end{align}
resulting in equilibrium natal BH spin [Eq.~\eqref{eq:aNS}]:
\begin{align}\label{eq:eqspin}
    &a_{\rm BH,eq} \approx 0.58\left(\frac{\RNS}{12\,{\rm km}}\right)^{1/2}\left(\frac{M}{3\,\msun}\right)^{-1/2}
    \times\notag\\&
    \Bigg[1+2.38\left(\frac{B}{10^{16}\,{\rm G}}\right)^{2}\left(\frac{\dot{M}}{\msun\,{\rm s}^{-1}}\right)^{-1}\left(\frac{\RNS}{12\,{\rm km}}\right)^{2}\Bigg]^{-1}\,,
\end{align}
where we have set $M = \mbh$. Eq.~\eqref{eq:eqspin} implies that the maximum equilibrium spin, as obtained by the PNS breakup velocity, is $a_{\rm BH,max} \approx 0.58\sqrt{R_{12}/M_3} $, where $ R_{12} = R/12\,\km $ and $ M_3 = M/3\,\msun $.
This result is consistent with full numerical solutions of rotating hydrostatic equilibrium \citep{Margalit2015}.

    \begin{figure*}
    \centering
        {\includegraphics[width=7.in]{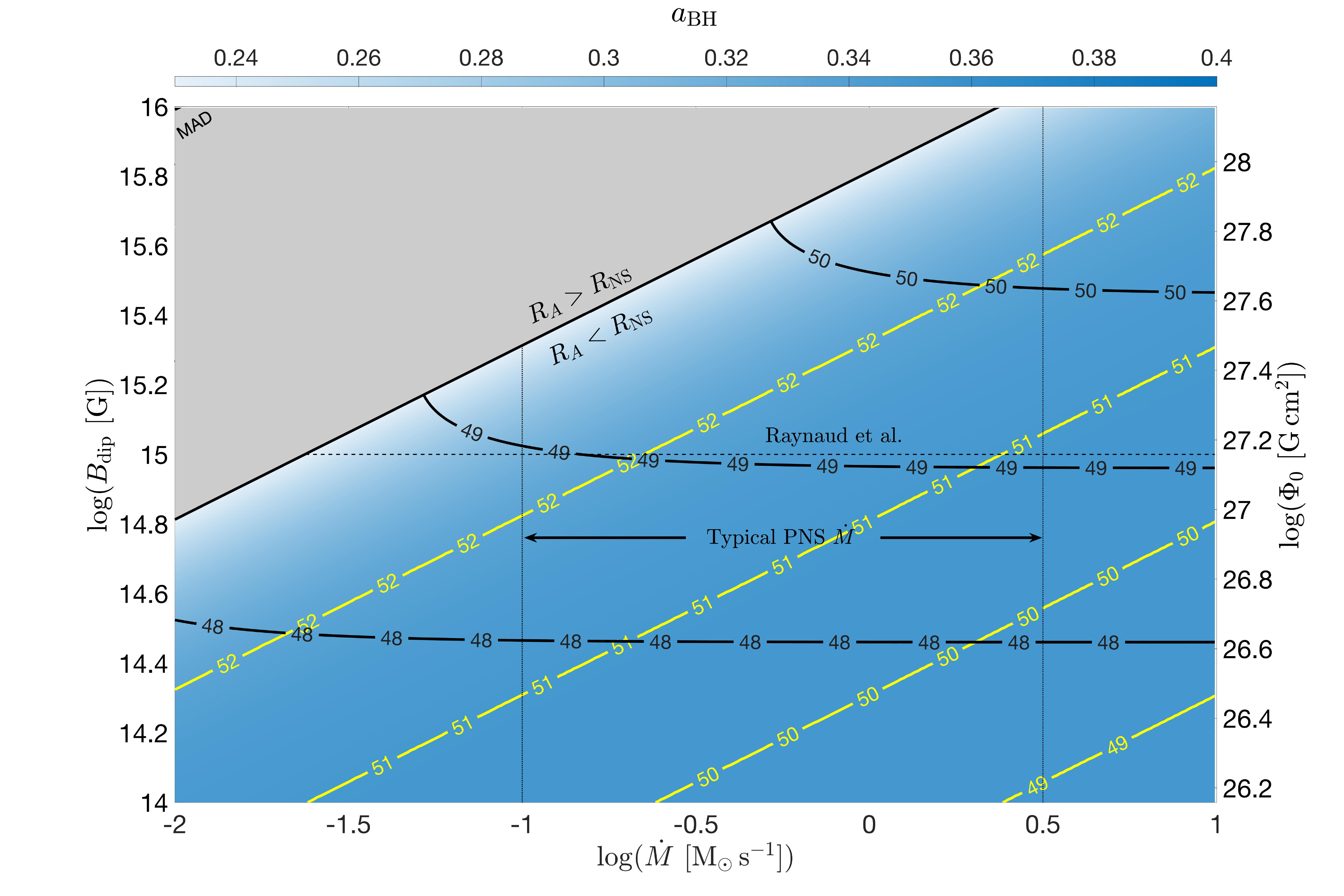}}
        \caption{
        Mapping of the PNS poloidal magnetic field strength (left axis), flux (right axis), and [assumed temporally constant] mass accretion rate onto the PNS (horizontal axis) to the natal BH spin (color map), PNS spin-down energy released prior to BH formation (yellow contours; in logarithmic units) and logarithmic BZ jet power (black contours), assuming initial $ P_0 = 1\,\ms $. The BH cannot inherit the PNS field if the Alfv\'{e}n radius fails to satisfy $ R_A \lesssim \RNS $ (the excluded area is in gray). The dotted vertical (dashed horizontal) black lines delineate the estimated range of mass accretion rates (PNS dipole field) at the time of collapse from PNS simulations by \citet{Aloy2021} \citep{Raynaud2020}. The moderate spin leads to jets with a characteristic GRB power. The MAD state with $ \eta_\phi \approx 1 $ is located deep in the gray zone, where the BH does not acquire the PNS field. The natal BH spin distribution does not change significantly for $ P_0 = 10\,\ms $.
        }
     \label{fig:collapse}
    \end{figure*}

    Figure~\ref{fig:collapse} shows key properties of the BH at the time of formation in the parameter space $ \{B,\dot{M}\}$, as determined by evolving $J(t)$ [Eq.~\eqref{eq:spin}] from $t=0$ to $t_{\rm BH}$ for $\RNS = 12\,\km $, $ M_0 = 1.7\,\msun $, $ \mbh = 2.7\,\msun $, $ I(t) = 0.35 M(t)\RNS^2 $, and initial $ P_0 = 1\,\ms $ (similar results are obtained for slower initial rotation, $ P_0 = 10\,\ms $). The color map reveals a rather narrow range of the BH spin at the time of formation ($\abh$), with values consistent with the inferred BH spins from GRB observations \citep{Gottlieb2023b}. The PNS spins down slower than $ \tBH $ such that the BH forms with a moderate spin throughout the parameter space (blue), i.e. we have $t_{\rm spin} \gg t_{\rm BH} \gtrsim t_B $. The PNS accretion rate range $ 0.1\,\msun\,\s^{-1} \lesssim \dot{M} \lesssim 3\,\msun\,\s^{-1}$ and the PNS dipole field \citep{Raynaud2020} are outlined as dotted and dashed black lines, respectively. The minimal condition for the BH to inherit the magnetic field of the PNS ($R_A < \RNS$) is satisfied below the thick black line, which covers a significant fraction of the parameter space.

    During the PNS spin-down phase, PNS winds are ejected into the collapsing star, inhibiting accretion along the polar axis and potentially preventing the magnetic field lines from being drawn into the BH upon collapse. The yellow contours delineate the logarithm of the spin-down energy emitted by the PNS before the collapse to a BH. Given the estimated accretion rate and dipole magnetic field of the PNS, the spin-down energy is $ \sim 10^{51}-10^{52}\,\erg $, and aligns with the excess energy observed in Type Ic-BL SNe associated with collapsar GRBs \citep[e.g.,][]{Cano2017}.

    The natal BH spin and magnetic flux determine the initial power of
    the BZ jet launched by the BH. Using Eq.~\eqref{eq:Pj}, we map the
    field and mass accretion rate to BZ-jet power upon collapse. The
    black contours illustrate that the $\log_{10}(P_\mathrm{j})$
    falls within the GRB luminosities inferred from observations
    \citep{Butler2010,Wanderman2010}. This implies that if
    $ \tBH \gtrsim \tB $, GRB jets likely emerge as soon as the BH forms.

In the opposite regime of $ \tBH \lesssim \tB $, there might not be enough time for the field to reach saturation before the PNS collapses. In such cases, the BH may form before an appreciable spin-down of the PNS. As a result, the BH will likely acquire a dimensionless spin similar to $ P_0 $. We conclude that for the expected $ 1\,\ms \lesssim P_0 \lesssim 10\,\ms $, the natal BH spin is moderate, irrespective of $ \tB/\tBH $, and the acquired field is less than the saturation level but may still reach high values depending on $ \tB/\tBH $. If $ \tB \lesssim \tBH $, the nascent BH forms with sufficient magnetic flux to launch a GRB-like jet immediately.

\section{Black hole magnetic field evolution}\label{sec:grmhd}

\subsection{GRMHD setup}\label{sec:setup}

We conduct a suite of GRMHD simulations utilizing the 3D GPU-accelerated code \textsc{h-amr} \citep{Liska2022}, leveraging an ideal equation of state featuring an adiabatic index of $ 4/3 $. This equation of state is well-suited for modeling the radiation-dominated gas prevalent in the core.  The simulations consider BHs embedded in a stellar envelope to investigate the evolution of the remnant magnetic field on the BH as a function of the BH spin, magnetic field, and presence of a disk.

In \S\ref{sec:bh}, we established that the BH likely forms with a moderate spin. However, depending on the value of $ \eta_\phi $, it may either spin down or spin up at later times. Consequently, we vary the dimensionless spin parameter across different simulation models. We emphasize that our simulations assume a static metric, maintaining a fixed BH spin throughout. A non-negligible spin-down within the simulation time $ T_s \gtrsim 6\,\s $ is expected if the initial BH spin is close to unity \citep{Jacquemin-Ide2024}. This implies that the simulated jets powered by rapidly spinning BHs will be somewhat stronger than if the spin-down effect were accounted for.

As our simulations do not model the PNS phase and consistently feature a central BH engine, we assume that the magnetic field of the collapsing PNS threads the BH upon formation, as outlined in \S\ref{sec:bh}. To initiate the BH with a strong magnetic field, we initialize the BH vicinity with a vertical magnetic field and total magnetic flux equivalent to that of the PNS. This approach facilitates early accumulation of the magnetic flux on the BH, with subsequent field evolution governed by the magnetohydrodynamics in the BH vicinity. Specifically, at the onset of the simulation, the core hosts a constant poloidal field of $ B_0 = 10^{12.5}\,{\rm G} $. The radial extent of the core magnetic field, determined by flux freezing, is given by $ R_c = \RNS\left(B_{\rm dip}/B_0\right)^{1/2} $. Despite initializing the flux with $ \Phi_0 \gtrsim 10^{28}\,{\rm G\,\cm^2} $, stochastic processes lead to a reduced flux on the BH, $ \Phi_s \lesssim 0.3\Phi_0 $, depending on the specific setup. Table~\ref{tab:models} outlines the different simulations considered here.

\begin{table}[]
    \setlength{\tabcolsep}{3.8pt}
    \centering
    \renewcommand{\arraystretch}{1.2}
    \begin{tabular}{| c | c c c c c | }

            \hline
        Model & $ \abh $ & $ \Phi_0~[10^{28}\,{\rm G\,\cm^2}] $ & $ \Phi_s~[10^{28}\,{\rm G\,\cm^2}] $ & Disk & $ T_s\,[\rm s] $\\	\hline
        $ a9\Phi hD $      & 0.9 & 3 & 1.0 & Yes & 6.3 \\
        $ a9\Phi hI $      & 0.9 & 3 & 0.3 & No  & 6.5 \\
        $ a1\Phi hD $      & 0.1 & 3 & 1.0 & Yes & 6.2 \\
        $ a9\Phi lD $      & 0.9 & 1 & 0.3 & Yes & 6.0 \\
        $ a5\Phi lD $      & 0.5 & 1 & 0.3 & Yes & 6.0 \\
        $ a1\Phi lD $      & 0.1 & 1 & 0.1 & Yes & 6.9 \\
            \hline
    \end{tabular}

    \caption{
        A summary of the models' parameters. The model names stand for the system type: the BH spin ($a$), total radial magnetic flux ($ \Phi $): high ($ h $) or low ($ l $), and disk presence ($ D $) or isolated compact object ($ I $). $ \abh $ is the BH spin, $ \Phi_0 $ is the total magnetic flux available for accretion onto the BH at the onset of the simulation, and $ \Phi_s $ is the flux retained by the BH after the initial phase. $ T_s $ is the total duration of the simulation.
        }
        \label{tab:models}

\end{table}

In the top panel of Fig.~\ref{fig:CC_profile}, the innermost shells initially contain excessive angular momentum to directly collapse into a BH. Namely, the gas within these shells has a circularization radius larger than the ISCO, indicating the formation of an accretion disk prior to BH formation. Viscosity in the disk will cause the gas to lose angular momentum, facilitating accretion onto the PNS and eventually leading to BH formation. Hence, the accretion disk is likely already established by the time of BH formation. To account for this process, we implement a rapid formation of an accretion disk following the angular momentum profile outlined in \citet{Gottlieb2022a,Gottlieb2022d}.

We set $ \mbh = 2.7\,\msun $ as an estimate of the maximum PNS mass, which varies with rotation and equation of state \citep[see discussion in][]{Margalit2017}. Our simulations do not include the core collapse and the PNS phase, which will affect the gas distribution in the collapsing star. Therefore, we initialize a spherically symmetric density profile by semi-analytically evolving the gas free-fall of the progenitor presented in \S\ref{sec:remnant} from the onset of core collapse to the BH formation time, $ \sim 1\,\s $ post-core collapse (see vertical dashed black line in Fig.~\ref{fig:CC_profile}). Hence, at the beginning of our simulations, the density profiles adhere to free-fall profiles with a power-law index of $ -1.5 $. We neglect the gas thermal pressure to compensate for the absence of self-gravity in the simulations.

For the grid setup, we employ spherical polar coordinates, $r$, $\theta$, $\varphi$. The $ r $-direction is logarithmically distributed, from slightly inside the event horizon to $ 4\times 10^4\,\km $. Uniform cell size distribution is maintained along the $\theta$- and $\varphi$-directions. The basic grid encompasses $N_r\times N_\theta \times N_\varphi $ cells in the $r$-, $\theta$-, and $\varphi$-directions, respectively, where $ N_r = 256 $; $ N_\theta = 128 $; and $ N_\varphi = 128 $. We incorporate local adaptive time-stepping and static mesh refinement to enhance computational efficiency. For all simulations, refinement is implemented at one level within the innermost $ r = 80\,\rg$ to ensure proper resolution of the wavelength of the fastest-growing MRI mode.

\subsection{Results}\label{sec:results}

    \begin{figure*}
    \centering
    \href{https://oregottlieb.com/videos/a9PhD.mp4}
        {\includegraphics[width=2.8in]{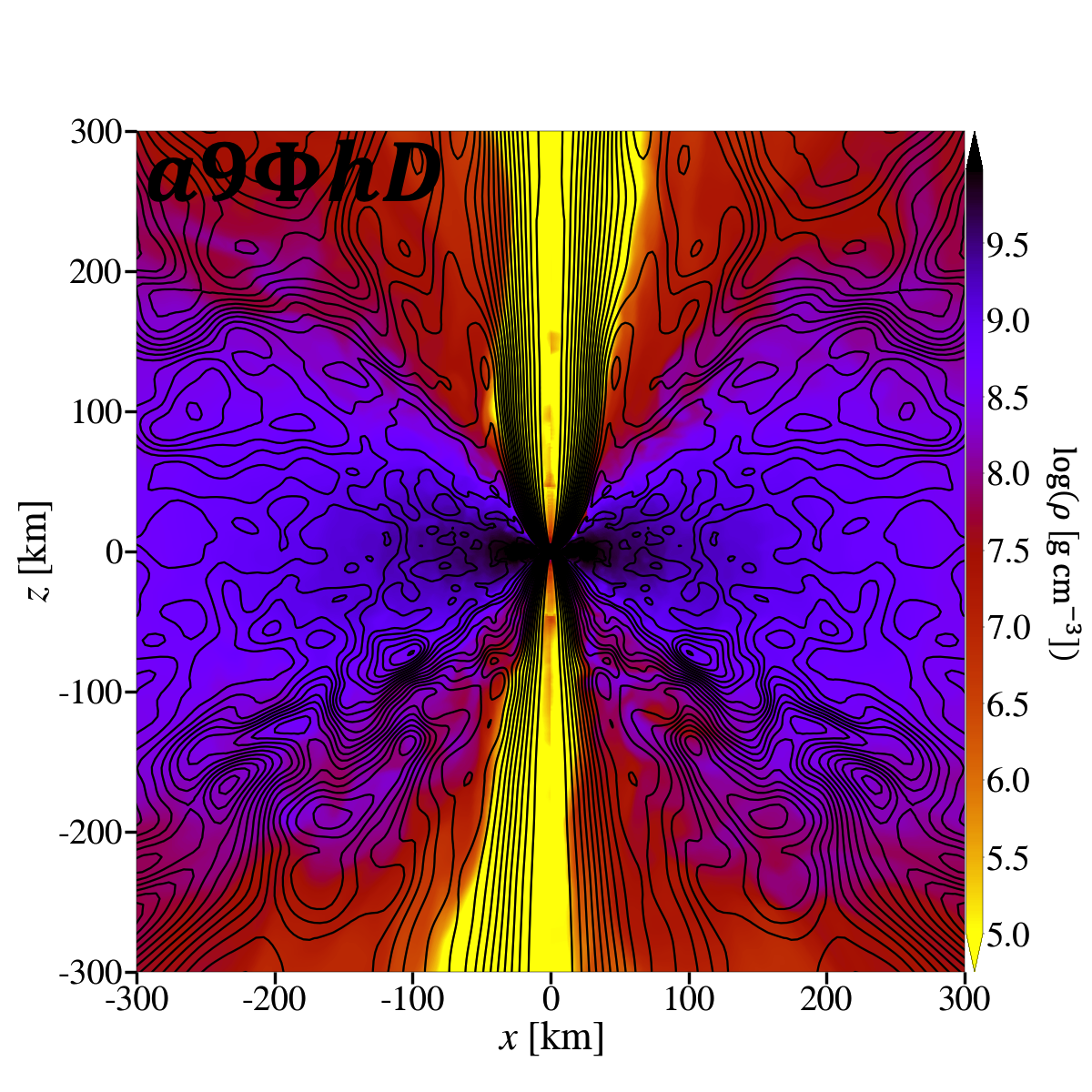}}
    \href{https://oregottlieb.com/videos/a9PhI.mp4}
        {\includegraphics[width=2.8in]{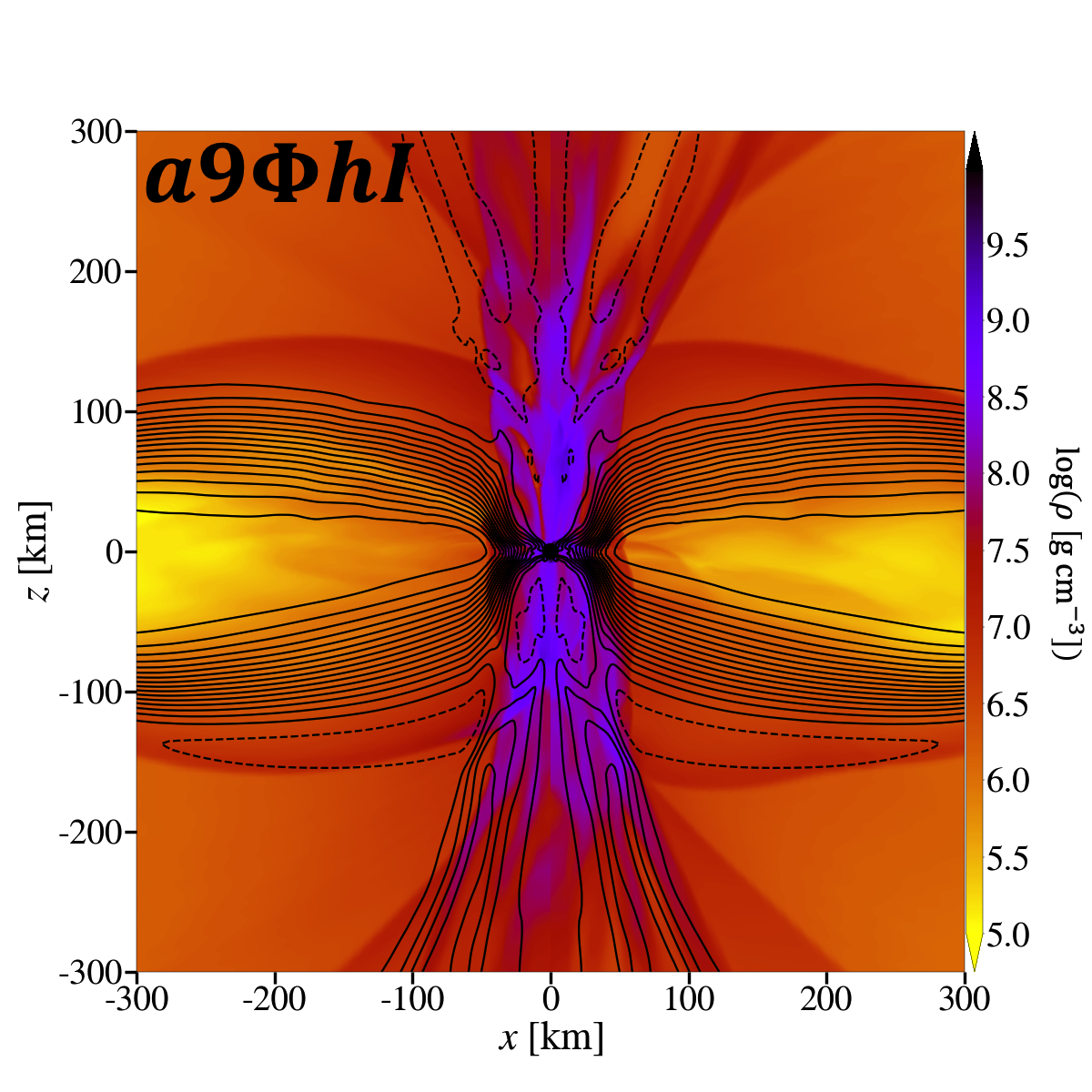}}
    \href{https://oregottlieb.com/videos/a1PhD.mp4}
        {\includegraphics[width=2.8in]{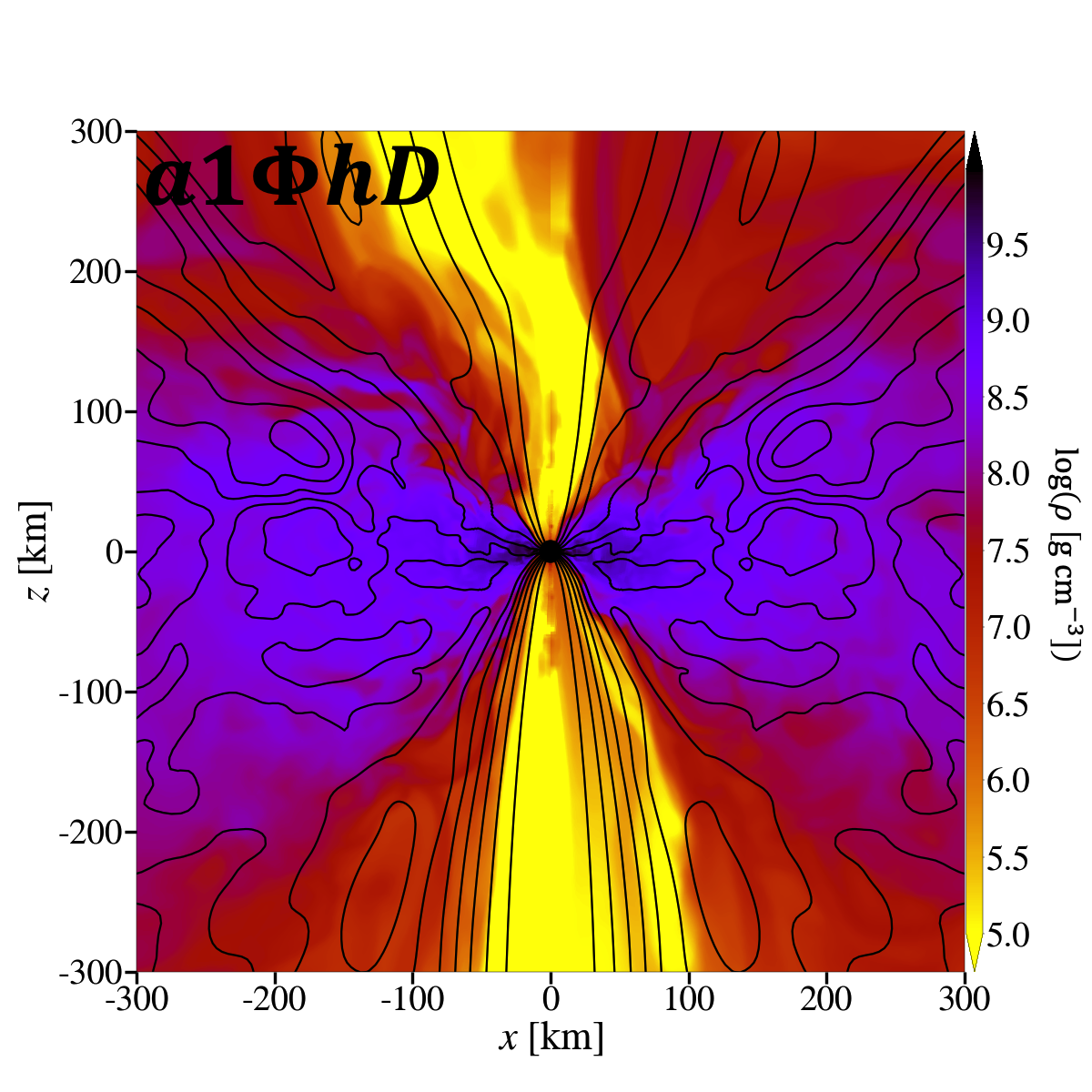}}
    \href{https://oregottlieb.com/videos/a9PlD.mp4}
        {\includegraphics[width=2.8in]{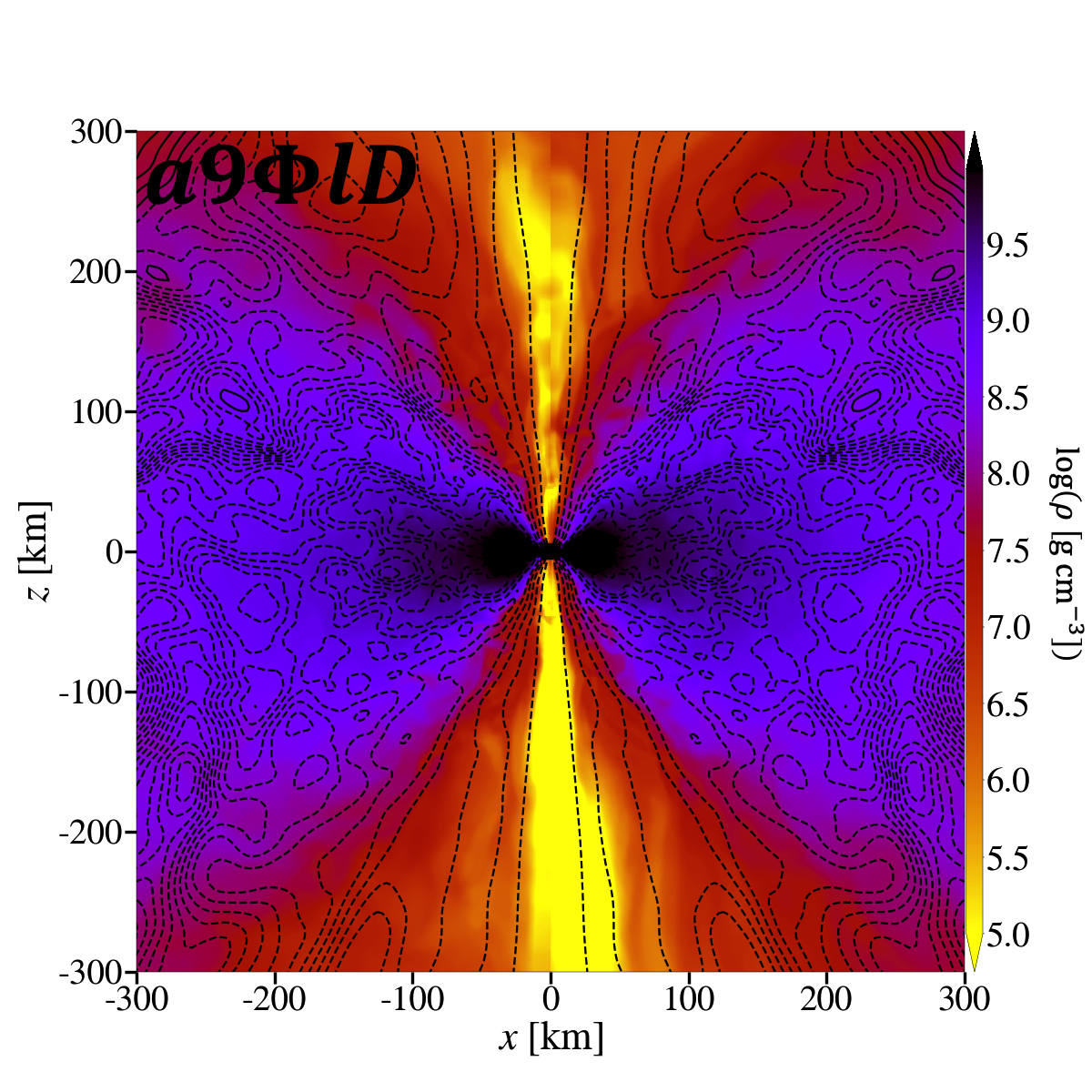}}
    \href{https://oregottlieb.com/videos/a5PlD.mp4}
        {\includegraphics[width=2.8in]{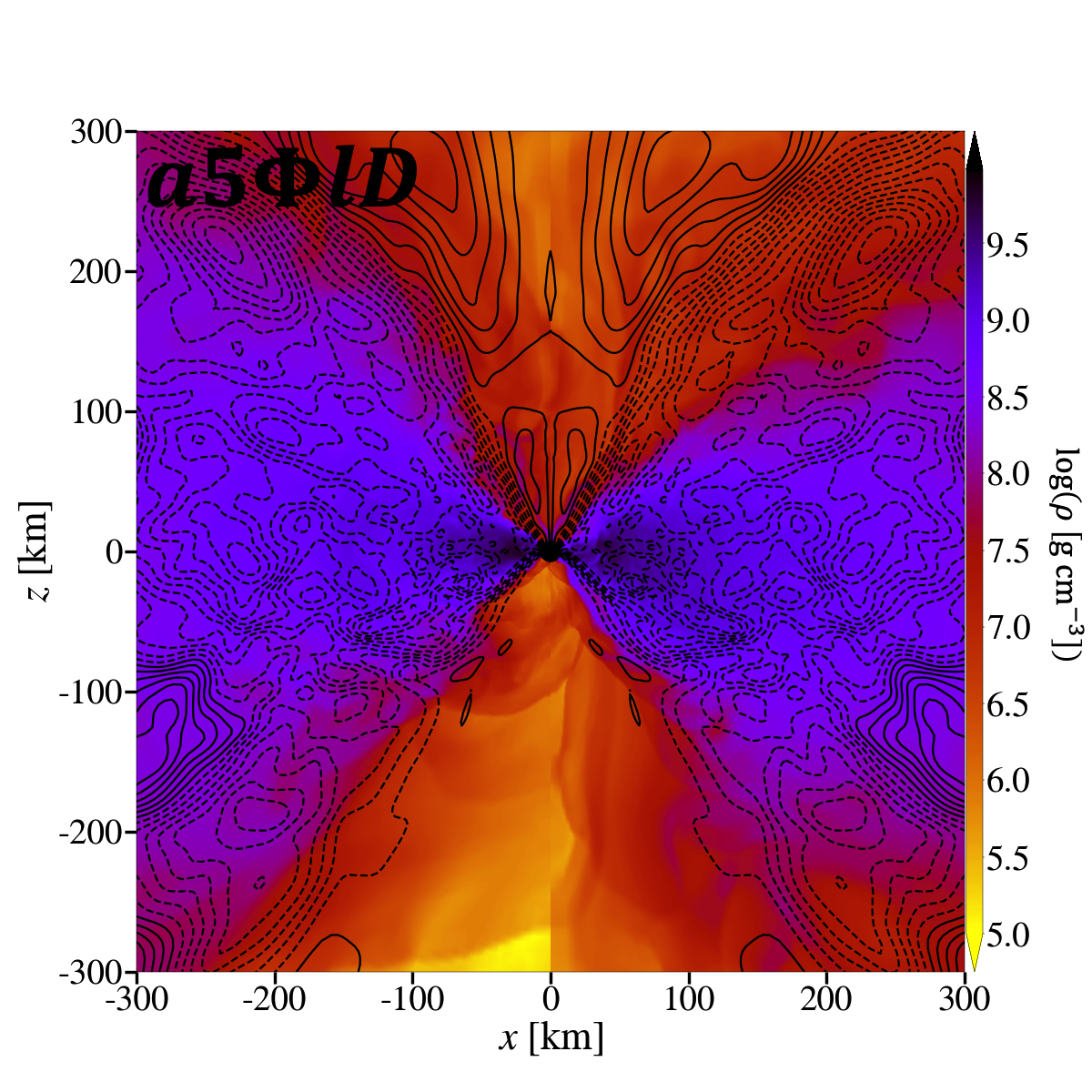}}
    \href{https://oregottlieb.com/videos/a1PlD.mp4}
        {\includegraphics[width=2.8in]{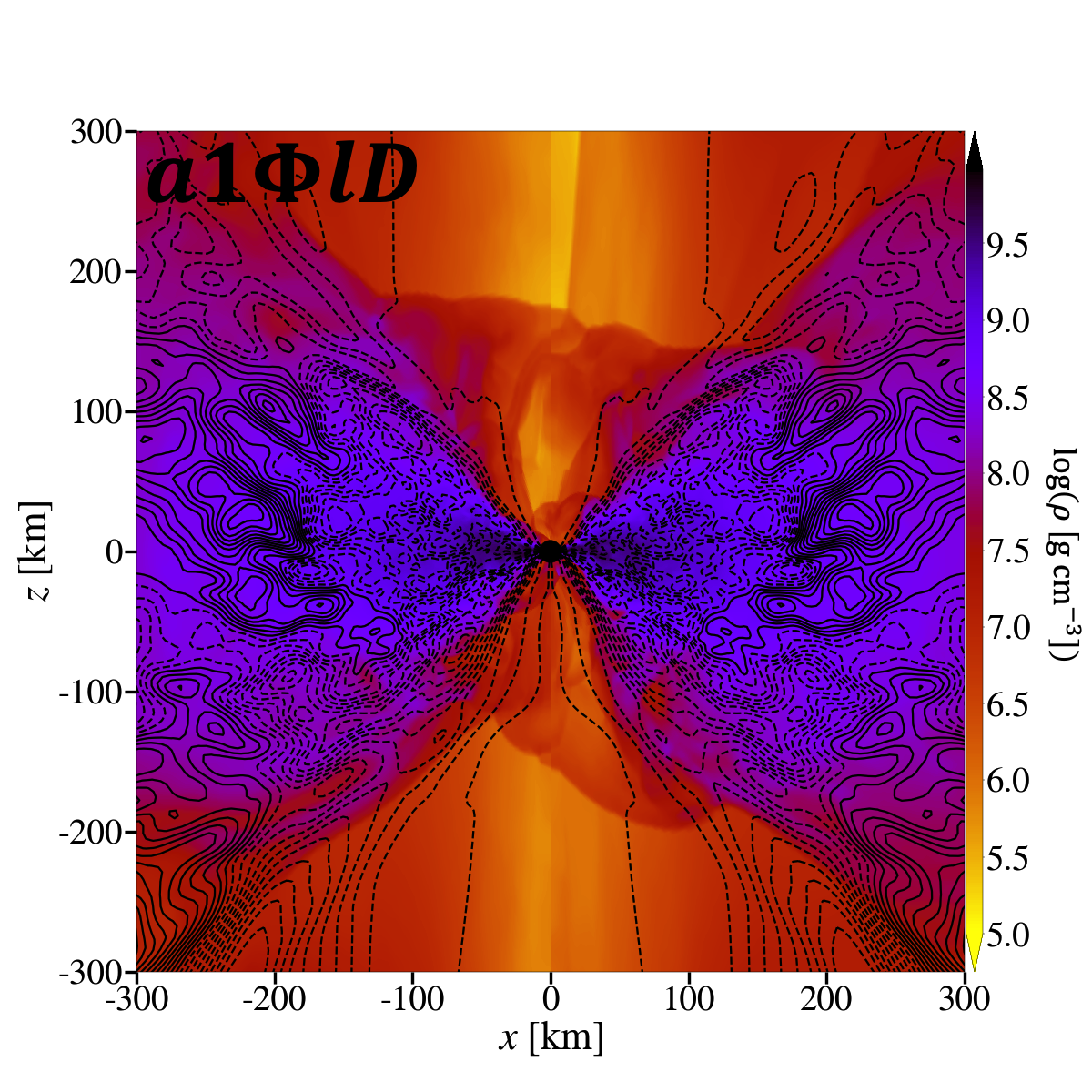}}
        \caption{Vertical slices of the logarithmic mass density in the simulated models, as measured at $ t = T_s $. The black contours outline the magnetic field lines, where solid and dashed lines delineate positive and negative magnetic polarity, respectively. BHs with higher spins and magnetic fluxes retain the remnant poloidal field of the PNS to maintain steady jet launching. When the BH spin and the initial flux on the BH are low, or an accretion disk is absent (model $ a9\Phi hI $), the BH cannot sustain a large-scale field. As a result, magnetic loops of opposite polarity are accreted onto the BH, causing the BH magnetic flux to reconnect and reduce the flux on the BH, resulting in the depletion of jets. Full movies showcasing the evolution of the magnetic field on the BH in each simulation are available at \url{http://www.oregottlieb.com/BH_field.html}.}
     \label{fig:maps}
    \end{figure*}

    \begin{figure*}
    \centering
        {\includegraphics[width=6.5in]{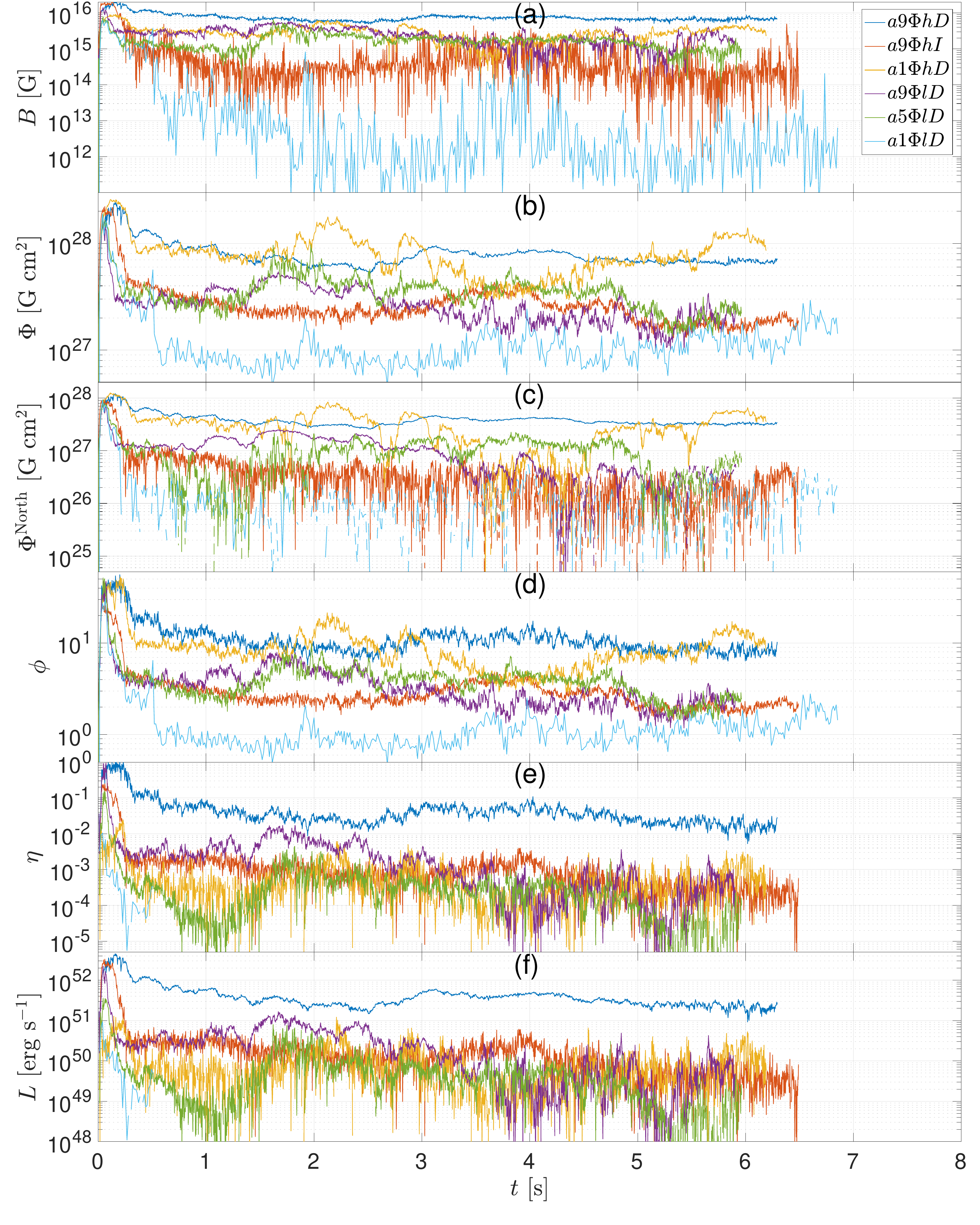}}
        \caption{Time evolution of physical quantities on the horizon.
        The magnetic field strength on the north pole {\bf (a)} and integrated flux {\bf (b)} illustrate that the BH retains high magnetization/flux when the initial magnetic field is strong or the BH spin is moderate or high.
        {\bf (c)}: the flux on the northern hemisphere indicates that in models in which magnetic flux reconnects away, negative polarity loops thread the horizon (dashed lines). The dimensionless flux {\bf (d)}, electromagnetic launching efficiency {\bf (e)}, and jet luminosity {\bf (f)} of models with initially strong magnetic field or moderate/high BH spin, produce jets. Otherwise, if the BH spin and initial field are too low, or there is no disk to confine the jet, no outflows are observed.
        }
     \label{fig:phi}
    \end{figure*}


As shown in \S\ref{sec:remnant}, the excess angular momentum in the core implies that the accretion disk will be present during the PNS collapse. Therefore, we simulate the collapse of a star with a rotation profile that facilitates the rapid formation of an accretion disk, before the total magnetic flux accumulates on the BH. In simulations with higher (lower) magnetic flux, most of the initial flux is advected onto the BH within $ \sim 0.2~(0.1)\,\s $. Over the following $ \sim 0.1\,\s $, some magnetic flux is lost to reconnection and stabilizes with $ \Phi_s \approx 0.3\Phi_0 $ threading the BH, if a disk is present. If there is no disk to hamper reconnection, the flux drops to $ \Phi_s \lesssim 0.1\Phi_0 $.

Figure~\ref{fig:maps} depicts vertical cuts of logarithmic mass density with contours of the magnetic field lines at $ t = T_s \gtrsim 6\,\s $ post-PNS collapse. Model $ a9\Phi hD $ exhibits a robust, steady, and collimated magnetized outflow powered by the BH with ordered poloidal magnetic field lines threading the BH. This model features a rapidly spinning BH surrounded by an accretion disk that confines the magnetic field on the BH by preventing loops of opposite polarity from reconnecting, thereby enabling the field to remain attached to the BH. While the BH launches near-MAD jets, the disk remains weakly magnetized throughout the simulation, illustrating minimum leakage of the field from the BH to the disk.

A relativistic jet is also observed in the models in the second row, where there is either a rapidly spinning BH with a lower initial flux ($ a9\Phi lD $) or a slowly spinning BH with a higher initial flux ($ a1\Phi hD $). However, in these configurations, the jet exhibits diminished strength attributed to the reduced field intensity or the slower rotation of the BH, compared to model $ a9\Phi hD $. This is demonstrated by the field line density along the relativistic jets in Fig.~\ref{fig:maps}. As a result, the jet manifests intermittency, with instances in which no jet is observed emanating from the BH, or only a one-sided jet is launched.

If both the BH spin and the magnetic flux threading the BH are low (bottom panels), the jet power is insufficient to clear the high-density plasma surrounding the BH. Over time, as more unmagnetized gas is accreted, the magnetic flux on the BH drops by virtue of reconnection (see plasmoids in the map of model $ a5\Phi lD $), and the jet launching process cannot be maintained. Thus, such configurations can support jet launching for some time, depending on the specific spin and flux. While such short-lived jets may unbind the stellar envelope and power a transient, they cannot generate typical collapsar GRBs that last for $ t \gtrsim 10\,\s $. Nevertheless, we stress that while our simulations do not include spin evolution, in reality, BHs in such configurations likely spin up, allowing more efficient jet launching at later times.

Finally, we examine the scenario of a high BH spin and strong field in the absence of an accretion disk ($ a9\Phi hI $). Although this situation is not typically expected as both the BH spin and the disk stem from the PNS rotation, we use it to demonstrate that in the absence of a disk, there is no structure to anchor the field on the horizon. Consequently, the field leaks out through magnetic reconnection, detaining from the BH. This results in a more spherical accretion onto the BH, causing an accumulation of gas around it. The map of $ a9\Phi hI $ reveals a distinct pattern compared to other models: high-density gas accumulates along the poles as the dipole field diminishes, whereas low-density gas congregates at the equator, owing to shielding of the magnetic field lines stretched along the equatorial plane. We conclude that, even with high BH spin and strong initial magnetic flux, the presence of an accretion disk is crucial for sustaining high flux on the BH and launching relativistic jets.

Figure~\ref{fig:phi} illustrates the temporal evolution of various parameters on the horizon. The BH maintains a strong magnetic field (a) and high aggregated flux (b) so long as an accretion disk is present, irrespective of the BH spin. However, models with low spin and low flux ($a1\Phi lD $) or without a disk struggle to sustain strong fields over long times. In Fig.~\ref{fig:phi}(c), which depicts the flux on the northern hemisphere, it is apparent that models with weak fields have their magnetic flux reconnect, forming opposite polarity loops threading through the horizon (dashed lines depict negative polarity). Even in the absence of the disk, some magnetic flux remains due to the quasi-spherical accretion that retains it, although it lacks a distinct dipolar field to launch jets.

The dimensionless magnetic flux (d) reveals that the flux is preserved in models with initially high flux and where an accretion disk is present. None of the models exhibit a MAD state ($ \phi < \phi_{\rm MAD} $), which is obtained only for very low mass accretion rates and strong magnetic fields (Fig.~\ref{fig:collapse}). However, a MAD state might be achieved as the mass accretion rate will ultimately go down. The spin efficiency plays a crucial role in the total electromagnetic launching efficiency (e) and jet luminosity (f). When both the flux and the spin are high ($ a9\Phi hD $), the jet power surpasses that of the observed GRB population \citep[see][]{Gottlieb2023b}. Therefore, as suggested in \S\ref{sec:bh}, the BHs that power GRB jets should have either a lower magnetic flux or spin, $ \abh \lesssim 0.5 $. Indeed, models featuring initially lower magnetic field strengths or moderate BH spins yield jets with power consistent with observations. Conversely, insufficient BH spin or initial field strength or the absence of a confining disk fail to produce relativistic outflows.

\section{Discussion}
\label{sec:discussion}

GRBs are believed to originate from magnetized BHs via the BZ process. Current models implicitly rely on one of two assumptions. One scenario is a magnetic field amplification through dynamo action in the accretion disk formed by infalling gas. This process has been numerically demonstrated in the context of post-merger disks \citep[e.g.,][]{Hayashi2022,Hayashi2023,Gottlieb2023d}. The dynamo process in the disk is efficient when the disk's radial extent is substantial \citep{Jacquemin-Ide+23} and the advection of the field is effective when the disk is not too thin \citep{Jacquemin-Ide2021}. Both collapsars and post-merger disks exhibit relatively compact and high mass density disks, leading to neutrino emission that thins out the disk \citep{Narayan2001,Kohri2005,Chen2007}. However, post-merger disks benefit from a sharp drop in the mass accretion rate \citep{Gottlieb2023e}, which halts the cooling and expedites the transition to MAD, allowing the magnetic flux on the BH to become dynamically important early on. In contrast, collapsars experience a continuous mass supply, keeping their disks extremely dense for extended periods. Our MESA models suggest that this likely prevents the disk from transitioning to MAD within the time of the collapse, precluding jet launching \citep{Gottlieb2022a}. However, further studies are required to investigate the dynamo process in disks obtained from stellar evolution models. Additionally, it is necessary to explore the range of pre-collapse structures as a function of the angular momentum transport \citep{Spruit2002,Fuller2019a,Skoutnev2024} and loss mechanisms \citep[e.g.,][]{vink:15, georgy:17}, initial rotation distributions \citep[e.g.,][]{ramirez-agudelo:15, britavskiy:24}, and the impact of binary interactions \citep[e.g.,][]{ Cantiello:2007, demink:13,renzo:23}, as well as other physical processes relevant to collapsar progenitors.

Alternatively, the strong magnetic field threading the BH could be advected during the free-fall collapse while maintaining flux freezing. We demonstrate that under the TSD mechanism, stellar evolution models suggest the magnetic field loops are too small to add up coherently. The resulting flux on the BH falls short by orders of magnitude compared to that needed to power a GRB jet. The inability of both scenarios to generate a strong large-scale poloidal magnetic field poses significant challenges in understanding the origin of the BH's magnetic field.

Our stellar evolution models reveal that rapidly rotating progenitors associated with GRBs exhibit a large amount of angular momentum in the core, forming a PNS with an accretion disk prior to BH formation. During the PNS phase, the convective dynamo within the PNS amplifies its internal magnetic field to $ B \approx 10^{15}\,\G $. As the PNS reaches a critical mass and collapses into a BH, some of its magnetic field is inherited by the BH, depending on the relative length of the balding timescale and the viscous time. We find that, for a characteristic mass accretion rate and PNS field, the viscous time is shorter than the balding timescale, allowing the field lines to rearrange and enabling the nascent BH to retain the flux. Conversely, if the mass accretion rate onto the PNS is very low and the magnetic field is extremely strong, the magnetic footprints lie outside the PNS, resulting in most of the flux being lost to reconnection during the collapse. Nonetheless, even if the balding time is too short, causing the bulk of the PNS flux to be lost during the collapse, the BH is likely to inherit the highly magnetized PNS accretion disk. This disk can serve as a seed for further amplification of the magnetic field, ultimately leading to the launch of a BZ jet. This suggests that even if the BH flux is generated by an accretion disk, it is likely that the disk is an inherited, highly magnetized PNS disk. Namely, there is no necessity for a newly formed accretion disk to gradually amplify the weaker fossil star fields through dynamo processes.

The natal properties of the magnetized BH depend on the available time for the PNS to amplify its magnetic field and evolve its spin before collapsing into the BH. Current numerical calculations estimate the PNS lifetime to be comparable to the amplification timescale, suggesting that the PNS has sufficient time to develop high dipolar fields of $ B \approx 10^{15}\,\G $. During this period, the PNS may either spin up due to accretion or spin down due to magnetic braking. Our findings indicate that at the time of collapse, the PNS is likely to end up with a moderate dimensionless spin parameter of $ \abh \approx 0.35 $. These parameters enable the BH to launch relativistic jets as soon as it forms. The corresponding BZ power for this field and spin is $ P_j \approx 10^{49}\,\erg\,\s^{-1} $.

To determine whether the BH can sustain the launch of jets without an additional supply of magnetic flux, we conducted 3D GRMHD simulations of BHs with the expected magnetic flux from the PNS embedded inside a star based on our stellar evolution models. Our findings indicate that an accretion disk is essential for the BH to retain its magnetic field, enabling a steady launch of relativistic jets throughout the simulations ($ t > 6\,\s $) with a roughly constant luminosity. Our jets are not in a MAD state, which requires higher magnetic flux and a lower mass accretion rate than current simulations predict. If no additional magnetic flux is accreted onto the BH, the BH will remain in a sub-MAD state. However, over time, the mass accretion rate is expected to decrease, increasing the dimensionless magnetic flux on the BH until it goes MAD. In scenarios where the disk is absent, the magnetic flux reconnects along the equator, preventing the BH from launching jets. Numerical relativity simulations that follow the PNS phase, the PNS-disk interaction, and the PNS collapse into a BH are crucial to provide a rigorous, self-consistent test of this scenario and will be conducted in a follow-up study.

\section{Big Picture Insights on gamma-ray bursts and associated supernovae}\label{sec:implications}

In this \emph{letter}, we have explored rapidly rotating stars that support the necessary conditions for accretion disk formation, which is essential for GRB jets, and subsequently lead to the birth of rapidly spinning magnetars. Such proto-magnetars typically collapse into BHs within $ \sim 1\,\s $. During this time, they release $ \sim 10^{51}-10^{52}\,\erg $ into the collapsing star, potentially accounting for the excess energy observed in Type Ic-BL SNe associated with collapsar GRBs. However, in some cases, the magnetar winds may impede accretion sufficiently to prevent BH collapse. Consequently, the ms magnetar may remain active for an extended period, potentially powering Type I superluminous SNe \citep[e.g.,][]{Kasen2010,Woosley2010,Nicholl2017,Vurm2021}.

In contrast, more common stars may possess stronger large-scale magnetic fields. These fields could be of fossil origin or generated through dynamo action in early evolutionary phases \citep[e.g.][]{Donati2009,fuller:15,Cantiello:2016}. The presence of such large-scale fields facilitates efficient angular momentum transport away from the core. As a result, these stars are expected to form slowly spinning magnetars, which might be associated with the SN engine.

Following the collapse of the PNS into a BH, a BZ driven jet with a constant power of $ P_j \sim 10^{49}\,\erg $ is launched into the collapsing star. Once this jet breaks out of the optically thick star, it can generate the prompt GRB emission. Both the constant jet power and its magnitude align with observations of typical GRBs \citep[e.g.,][]{Mcbreen2002,Butler2010}, where the GRB rapid variability naturally emerges from the jet-star interaction \citep{Gottlieb2019}. Nonetheless, GRBs exhibit a wide spectrum of luminosities \citep[e.g.,][]{Wanderman2010}. Within our model, the jet power varies with the BH spin and the magnetic flux. Namely, the range of GRB luminosities may stem from variations in mass accretion rates and magnetic fluxes on the PNS, which in turn determine the resulting natal BH spin and dipolar field. During the jet activity, the BH is expected to enter the MAD state, leading to accelerated BH spin-down. Therefore, the end point of the jet launching is likely governed by either the decreasing mass accretion rate, the spin-down of the BH, or a combination of both factors [see Eq.~\eqref{eq:Pj}].

Finally, the current paradigm posits two stringent requirements for the progenitor stars of GRBs: they must simultaneously maintain strong magnetic fields and rapid rotation \citep[e.g.,][]{Symbalisty1984, Metzger2008, Gottlieb2022a}. However, strong magnetic fields carry away angular momentum from the star through magnetized winds \citep[e.g.,][]{Varma2023,Petitdemange2023,Petitdemange2024}, making the coexistence of a strong field and rapid rotation challenging.
If BHs inherit their magnetic field from PNSs, as proposed here, these requirements are alleviated, making rapid rotation of the star the primary requirement for jet launching. This suggests the following: (1) Relativistic jets might form more readily and, hence, be more common. Furthermore, it also suggests that the majority of collapsar BH accretion disks will be accompanied by magnetized jets; (2) BHs formed through direct collapse such as in pair-instability SNe cannot achieve strong magnetic fields and thus cannot produce jets, implying that a PNS phase is crucial for typical collapsar GRBs. Other GRB classes, such as ultra-long GRBs, may require a different launching mechanism, suggesting that they are fundamentally distinct from ordinary GRBs.

\acknowledgements

We are grateful to Lucy Reading-Ikkanda/Simons Foundation for designing Figure~\ref{fig:sketch}. We thank Valentin Skoutnev, Amir Levinson, Wenbin Lu, Jonatan Jacquemin-Ide, Elias Most, Jim Fuller, Stan Woosley, and Adam Burrows for their helpful comments. MR thanks Aldana Grichener for useful feedback on the reproducibility package on zenodo.
OG and JAG are supported by the Flatiron Research Fellowship. BDM was supported in part by the National Science Foundation (grant No. AST-2009255) and by the NASA Fermi Guest Investigator Program (grant No.~80NSSC22K1574). The Flatiron Institute is supported by the Simons Foundation. The computations in this work were, in part, run at facilities supported by the Scientific Computing Core at the Flatiron Institute, a division of the Simons Foundation.
This research used resources of the National Energy Research Scientific Computing Center, a DOE Office of Science User Facility supported by the Office of Science of the U.S. Department of Energy under Contract No. DE-AC02-05CH11231 using NERSC allocations m4603 (award NP-ERCAP0029085).

\bibliography{refs}

\appendix

\section{Evolution of the stellar progenitor}
\label{appendix:MESA}

We compute the stellar progenitor using MESA (version r24.03.1) with the input files available at \href{https://doi.org/10.5281/zenodo.12193630}{doi.org/10.5281/zenodo.12193630}. To obtain a reliable core structure (including free electron profile), we use the 128-isotope nuclear reaction network \texttt{mesa\_128.net} throughout the evolution. This is necessary to describe the weak reactions during and beyond silicon core burning \citep[][]{farmer:16}. We emphasize that with $\sim$20-isotope networks, the free electron profile of the core, and consequently the density and angular momentum profiles are demonstrably unreliable \citep{farmer:16, renzo:24RNAAS}. We adopt nuclear reaction rates that are a combination of NACRE \citep{Angulo1999}, JINA REACLIB \citep{Cyburt2010}, plus additional tabulated weak reaction rates \citet{Fuller1985, Oda1994, Langanke2000}. We include electron screening following \citet{Chugunov2007} and thermal neutrino losses following \citet{Itoh1996}.  In our setup, radiative opacities are primarily from OPAL \citep{Iglesias1993, Iglesias1996} and electron conduction opacities are from \citet{Cassisi2007}.

Throughout the evolution, we include the wind mass loss rate following \cite{vink:00} and \cite{nugis:00} assuming the metallicity scaling from \cite{vink:01} and the rotational enhancement from \cite{langer:98}. We include core overshooting following \cite{brott:11} and treat rotational mixing following \cite{heger:00} including angular momentum transport with a ``classical'' Tayler-Spruit dynamo \citep{Spruit2002}.

After the formation of a carbon-oxygen core, we prevent the development of spurious numerical velocities in the outer layers by artificially setting to zero the velocity in any layer that has a sound-propagation time from the outer edge of the core, or the location where the specific entropy drops below 10.5\,$N_{A}k_{B}$, whichever is further in, longer than the current timestep. Similar artificial damping exists in all calculations of post-core-carbon burning massive stars \citep[e.g.,][]{Aguilera-Dena2018}.

\begin{figure}[htbp]
  \centering
  \includegraphics[width=\textwidth]{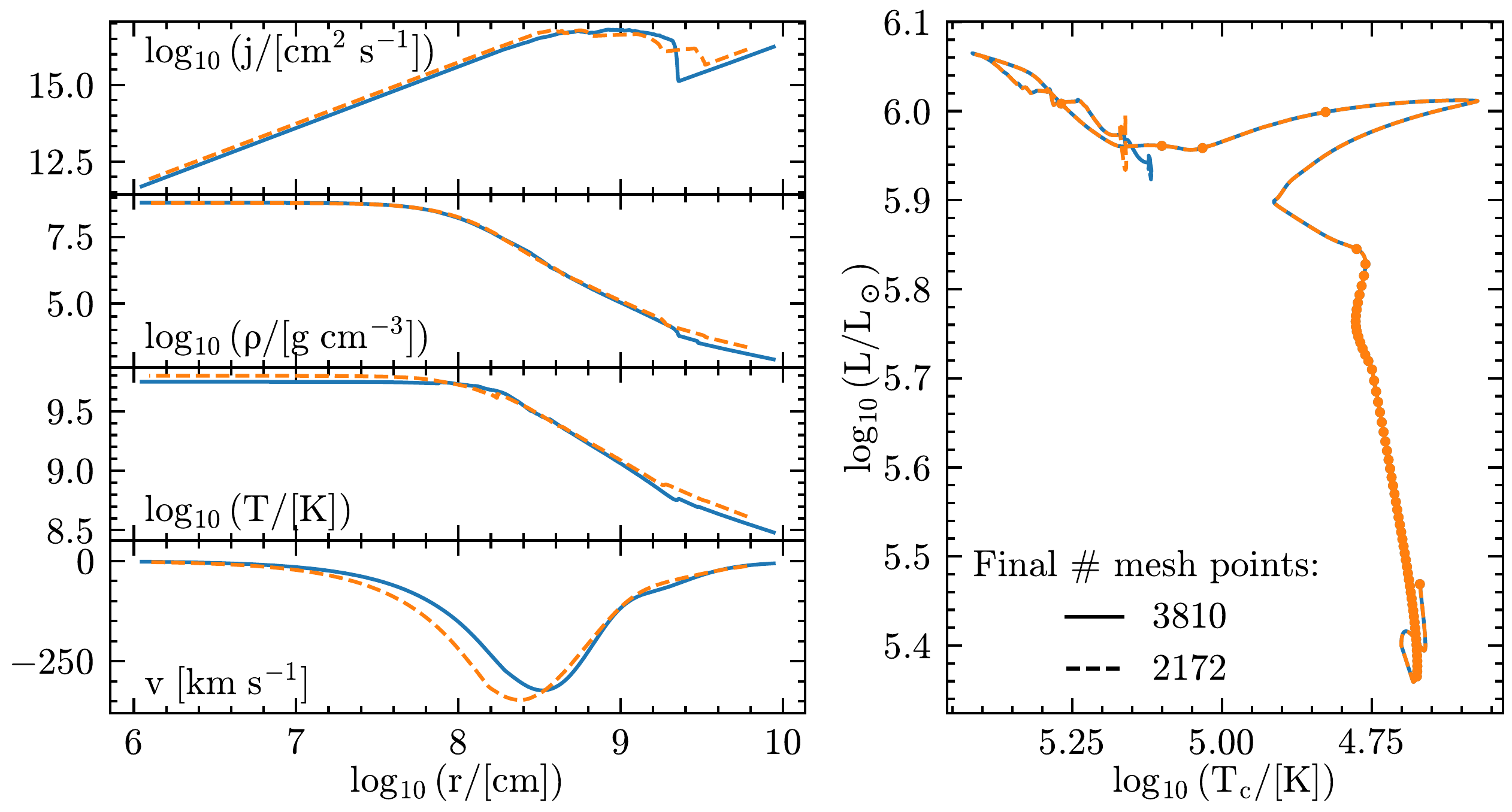}
  \caption{Resolution test for $40\,M_\odot$, $Z=0.001$ initially
    rotating at $0.6\omega_\mathrm{crit}$ model, which experiences
    rotationally induced chemically homogeneous evolution. The left panels
    show, from top to bottom, the specific angular momentum, density,
    temperature, and velocity profile as a function of radius at the
    onset of core-collapse. The right panel shows the
    Hertzsprung-Russell diagram, where each dot marks $10^5$\,years of
    evolution. The label indicates the number of mesh points at the
    onset of core collapse. The higher resolution model was recomputed
    post oxygen core depletion only.}
  \label{fig:res_test}
\end{figure}

The right panel of Figure~\ref{fig:res_test} shows the Hertzsprung-Russell diagram for this star, which evolves blueward and remains smaller than $25\,R_\odot$ throughout the evolution. At the end of the main sequence, roughly at 5.89\,Myr, it briefly contracts (analogously to the ``Henyey hook'' in non-rotating massive stars) and expands again at the ignition of Helium in the core. The remaining evolution occurs at roughly constant luminosity, set by the core mass, which corresponds to the total current mass because of strong rotational mixing.

The minimum radius $\sim{}0.5\,R_\odot$ is reached roughly at the onset of Neon core ignition and the evolution afterward is particularly sensitive to the treatment of convection, numerical resolution, and nuclear physics. To improve the numerical convergence, we increase the resolution in Lagrangian mass coordinates based on the radiative, adiabatic and mean molecular weight gradients across the local pressure scale height with a custom \texttt{other\_mesh\_fcn\_data} (see \texttt{run\_star\_extras.f90} available at \href{https://doi.org/10.5281/zenodo.12193630}{doi.org/10.5281/zenodo.12193630}).

We evolve the star until the onset of core-collapse, which, for our
purposes, we define as
$v_\mathrm{infall}\leq -300\,\mathrm{km\ s^{-1}}$. We verified that by
the time the iron core infall velocity exceeds this limit, all
processes of angular momentum transport and magnetic field generation
are effectively frozen for the remaining lifetime until the density
and temperature are such that the equation of state we use \citep[from
][]{skye} does not apply anymore. Therefore, this infall velocity
threshold is sufficient for our purposes.

The left panels of Fig.~\ref{fig:res_test} show a resolution test on the structure at the onset of core-collapse. We recomputed the model from oxygen core depletion (defined as the first time during the evolution after the end of hydrogen core burning when the central mass fraction of $^{16}\mathrm{O}$ is less than 0.1 \emph{and} the mass fractions of $^{12}\mathrm{C}$ and $^4\mathrm{He}$ are below 0.001 and 0.005, respectively) with increased spatial and temporal resolution (\texttt{mesh\_delta\_coeff},  \texttt{mesh\_delta\_coeff\_for\_highT}, and \texttt{mesh\_time\_coeff} set to 0.75, cf.~1.0 in our standard setup). This produces the dashed orange lines in Fig.~\ref{fig:res_test}. The legend in the right panel gives the number of mesh points in the profiles at the onset of core collapse. The blue model corresponds to the one discussed in the main text.

\end{document}